\newif\ifsubmode
\submodefalse

\ifsubmode
\documentclass[10pt,preprint]{aastex}
  \received{}
  \revised{}
  \accepted{}
\else
  \documentclass{emulateapj}   \usepackage{apjfonts}
\fi

\usepackage{epsf}
\usepackage{color}
\usepackage{amsmath}
\ifsubmode
  \newcommand{\authorspace}{\vspace{0pt}}
\else
  \newcommand{\authorspace}{\vspace{-10pt}}
\fi

\newcommand{\hst}{\textit{HST}}

\newcommand{\spitzer}{\textit{Spitzer}}

\newcommand{\ks}{\hbox{$K_s$}}

\newcommand{\ha}{\hbox{H$\alpha$}}
\newcommand{\paa}{\hbox{Pa$\alpha$}}
\newcommand{\lsim}{\lesssim}
\newcommand{\gsim}{\gtrsim}

\newcommand{\etal}{et al.}
\newcommand{\eg}{e.g.}

\newcommand{\cf}{cf.}

\newcommand{\msol}{\hbox{$M_\odot$}}

\newcommand{\zsol}{\hbox{$Z_\odot$}}

\newcommand{\lsol}{\hbox{$L_\odot$}}

\newcommand{\ujy}{\hbox{$\mu$Jy}}
\newcommand{\lir}{\hbox{$L_{\mathrm{IR}}$}}

\newcommand{\target}{SMM J163554.2+661225}
\shorttitle{PASCHEN--$\alpha$ EMISSION IN A GRAVITATIONALLY LENSED GALAXY}
\shortauthors{PAPOVICH ET AL.}

\begin{document}

%\slugcomment{\it Draft Version \today, \ampmtime}
\slugcomment{\it Draft Version \today}
\slugcomment{Accepted for Publication in the Astrophysical Journal}
%\slugcomment{Submitted to the Astrophysical Journal}
\title{PASCHEN--$\alpha$ EMISSION IN THE
GRAVITATIONALLY LENSED GALAXY SMM J163554.2+661225\altaffilmark{1}}

\author{\sc Casey Papovich\altaffilmark{2}}
\affil{George P.\ and Cynthia W.\ Mitchell Institute for Fundamental Physics and Astronomy, and \\
Department of Physics, Texas A\&M University, College Station, TX,
77843-4242}
\author{\sc \authorspace
Gregory Rudnick}
\affil{Department of Physics and Astronomy, 1251 Wescoe Hall Dr., University of Kansas, Lawrence, KS, 66045-7582}
\author{\sc\authorspace  Jane R.\ Rigby\altaffilmark{3}}
\affil{Observatories, Carnegie Institution of Washington, 813 Santa Barbara St., Pasadena, CA, 91101}
\author{\sc\authorspace Christopher N.~A.~Willmer}
\affil{Steward Observatory, University of Arizona, 933 N.\ Cherry
Ave., Tucson, AZ 85721}
\author{\sc\authorspace J.--D.~T.~Smith}
\affil{Ritter Observatory, University of Toledo, MS 113, Toledo, OH 43606}
\author{\sc\authorspace Steven L.~Finkelstein}
\affil{George P.\ and Cynthia W.\ Mitchell Institute for Fundamental Physics and Astronomy, and \\
Department of Physics, Texas A\&M University, College Station, TX,
77843-4242}
\and
\author{\sc Eiichi Egami and Marcia Rieke}
\affil{Steward Observatory, University of Arizona, 933 N.\ Cherry
Ave., Tucson, AZ 85721}
\altaffiltext{1}{ This work is based in part on observations made with the Spitzer Space Telescope, which is operated by the Jet Propulsion laboratory, California Institute of Technology, under NASA contract 1407.}
\altaffiltext{2}{papovich@physics.tamu.edu}
\altaffiltext{3}{Spitzer Fellow}

%%%%%%%%%%%%%%%%%%%%%%%%%%%%%%%%%%%%%%%%%%%%%%%%%%%%%%%%%%%%%%%%%%%%%%

\setcounter{footnote}{2}

\begin{abstract} We report the detection of the \paa\ emission line in
the $z=2.515$ galaxy \target\ using \spitzer\ spectroscopy.  \target\
is a sub-millimeter--selected infrared (IR)--luminous galaxy
maintaining a high star--formation rate (SFR), with no evidence of an
AGN from optical or infrared spectroscopy, nor X-ray emission.   This
galaxy is lensed gravitationally by the cluster Abell 2218, making it
accessible to \spitzer\ spectroscopy.  We measure a line luminosity,
$L(\paa)=(2.05\pm0.33)\times 10^{42}$~erg~s$^{-1}$, corrected for
gravitational lensing.  Comparing the \ha\ and \paa\ luminosities, we
derive a nebular extinction, $A(V)=3.6\pm0.4$~mag.  The
dust--corrected luminosity,
$L(\paa)=(2.57\pm0.43)\times10^{42}$~erg~s$^{-1}$, corresponds to an
ionization rate, $Q_0=(1.6\pm0.3)\times10^{55}$ $\gamma$ s$^{-1}$.
The instantaneous SFR is $\psi=171\pm28$~\msol~yr$^{-1}$, assuming a
Salpeter--like initial mass function from 0.1 to 100~\msol\ yr$^{-1}$.
The total IR luminosity derived using 70, 450, and 850~\micron\ data
is $\lir = (5-10)\times 10^{11}$~\lsol, corrected for gravitational
lensing.  This corresponds to $\psi=90-180$~\msol\ yr$^{-1}$, where
the upper range is consistent with that derived from the \paa\
luminosity.  While the $L(8\micron)/L(\paa)$ ratio is consistent with
the extrapolated relation observed in local galaxies and star--forming
regions,  the rest--frame 24~\micron\ luminosity is significantly
lower with respect to local galaxies of comparable \paa\ luminosity.
Thus,  \target\ arguably lacks a warmer dust component
($T_D\sim70$~K), which is associated with deeply embedded star
formation, and which contrasts with local galaxies with comparable
SFRs.  Rather, the starburst in \target\  is consistent with
star--forming local galaxies with intrinsic luminosities,
$\lir\approx10^{10}$~\lsol, but ``scaled--up'' by a factor of
$\sim$10--100.
\end{abstract}
 

%%%%%%%%%%%%%%%%%%%%%%%%%%%%%%%%%%%%%%%%%%%%%%%%%%%%%%%%%%%%%%%%%%%%%%

\section{INTRODUCTION}

There has been a growing industry of deep, multiwavelength surveys,
devoted to studying star formation and evolution of galaxies. Studies
of data from these surveys have concluded that the global
star--formation rate (SFR) density reached a maximum between
$z$$\sim$1.5--3 \citep[\eg,][and references therein]{hop04}, during
the period of rapid growth in the stellar mass density
\citep[\eg,][]{dic03,rud03,rud06}.  However, the study of SFRs in
these multiwavelength datasets requires estimates of the SFR using
available observables.  Common SFR indicators for high--redshift
galaxies  are  the rest--frame UV luminosity, especially at $z>1$ when
the UV shifts into the optical bands
\citep[\eg,][]{mad96,ste99,gia04,bou06,saw06,red08},  the far--IR
luminosity inferred from sub--mm emission \citep[\eg,][]{blain02}, the
mid--IR luminosity from \spitzer/24~\micron\
\citep[\eg,][]{per05,cap06,pap06,red06,webb06,dad07,franx08}, and the
$\ha$--emission line \citep[\eg,][]{erb06,bou07,kriek08}.     Each of
these SFR indicators has inherent uncertainties \citep[\eg,][and
references therein]{ken98,hop06}, and some that are unique to
observations at these redshifts (e.g., uncertainties in the
luminosity--temperature dependence on the sub-mm--\lir\ conversion,
Chapman et al.\ 2005; Pope et al.\ 2006; template incompleteness and
photometric redshift uncertainties on the 24~\micron--\lir\
conversion, Papovich et al.\ 2006, Reddy et al.\ 2006).
Nevertheless, several studies have compared SFRs derived from the UV,
24~\micron, far--IR, and sub--mm (and less--understood SFR tracers,
such as radio and X--ray emission), concluding broadly that while they
are consistent statistically, there is large scatter
\citep[\eg,][]{dad05,dad07,red06,pope06,pap07}.  There have been few
comparisons of direct, robust SFR indicators in individual galaxies at
high redshifts (e.g., Siana et al.\ 2008; references above). 

%However, even though these various tracers may be consistent with %one another statistically, there have been \textit{no} comparisons %against direct, robust SFR indicators. 

SFR indicators are normally calibrated against \ion{H}{1}
recombination lines \citep[\eg,][]{ken98}.  The primary advantage of
\ion{H}{1} recombination lines is that they effectively re-emit the
stellar luminosity from hydrogen--ionizing photons.  They are strong
in star--forming regions, and they trace directly emission by massive
stars.  The physics of \ion{H}{1}  emission in star--forming regions
is relatively well understood (Osterbrock 1989), and they exhibit only
a weak dependence on electron density and temperature.  Of the
\ion{H}{1} lines, \paa\ at $\lambda=1.8751$~\micron\ is an ideal
practical SFR indicator.   This line is exceptionally strong in
star--forming regions (1~\msol\ yr$^{-1}$ corresponds to $L[\paa] =
1.48 \times 10^{40}$ erg s$^{-1}$; Alonso--Herrero et al.\ 2006).
Furthermore, while recombination lines are subject to strong
attenuation, \paa\ suffers minimal extinction due to its longer
wavelength.  Recent studies of star formation in nearby galaxies and
star--forming regions within galaxies rely on the \paa\ luminosity to
interpret and to calibrate other SFR indicators, including the
infrared emission \citep[\eg,][]{cal05,cal07,alo06,ken07,dia08}.  For
extremely  dusty, star--forming IR luminous galaxies  with $A(V)
\sim$10--30~mag (e.g., Murphy et al.\ 2001; Dannerbauer et al.\ 2005;
Armus et al.\ 2007), the extinction at \paa\ is $\lsim $2~mag, and
optically thin.  Thus, even in the most obscured, luminous
star--forming regions, the \paa\ line traces the intrinsic ionization
rate, and thus the SFR.

While a few studies have compared SFR indicators in distant galaxies
to the \ha\ emission line (see references above),  no attempt to use
\paa\ as a SFR indicator has yet been attempted in any galaxy of
significant redshift.  At $z$$\gsim$2 the \paa\ $\lambda
1.875$~\micron\ line shifts to $\lambda_\mathrm{obs} > 5.2$~\micron,
and is accessible to the Infrared Spectrograph (IRS) on--board
\spitzer, which covers the wavelength range of 5.2--38~\micron.
However, star--forming galaxies at $z$$>$2 are too faint intrinsically
for observations with IRS at the expected wavelength for \paa\
(5.4--8~\micron), except for rare, extremely luminous objects, in
which AGN likely dominate the \ion{H}{1} line emission (e.g., Brand et
al.\ 2006), complicating any SFR--tracer comparisons.  We have
initiated a program using \spitzer/IRS to study the \paa\ emission in
12 galaxies at $z > 2$ gravitationally lensed by foreground galaxies
or clusters of galaxies.  Here, we discuss the detection of \paa\ in a
star--forming galaxy at $z=2.515$, gravitationally lensed by the rich cluster Abell~2218 \citep{kne04}.  This is the highest--redshift object with a measurement of
the \paa\ emission line yet obtained, and thus
we are able to study SFR indicators in a more ``typical'' star-forming
galaxy directly at high redshift.  

The outline for the rest of this paper is as follows.  In \S~2, we
summarize the properties of \target.  In \S~3, we discuss the
\spitzer\ observations and data reduction.   In \S~4, we discuss the
analysis of the spectroscopic and imaging data to derive rest--frame
luminosities.  In \S~5, we compare the rest--frame luminosities of
\target\ to local samples, and discuss the utility of the various
quantities as SFR indicators.  In \S~6, we present our conclusions.
In the Appendix, we use stellar population models to analyze the
rest--frame UV to near-IR spectral energy distribution (SED) of
\target, and make some conclusions on the nature of the stellar
populations in this galaxy.      To derive physical quantities we use
a cosmological model with $H_0=70$~km s$^{-1}$ Mpc$^{-1}$,
$\Omega_{M,0} = 0.3$, $\Omega_{\Lambda,0}=0.7$.    

\section{\target}

We targeted \target\ ($16^h35^m54.2^s$,
$+66^\circ12^\prime24.5^{\prime\prime}$, J2000), which is a galaxy at
$z=2.515$ gravitationally lensed by the galaxy cluster Abell~2218.
\target\ was identified by \citet[hereafter K04]{kne04} as a
lensed--galaxy candidate selected as a sub--mm source at 450 and
850~\micron, with a very red optical counterpart.  This source
consists of a triply lensed system, with a combined magnification from
gravitational lensing $\mu \sim 45$ (K04).    The brightest component
(component ``B'' in the notation of K04) has $\mu = 22\pm 2$, $K_s =
19.5$~mag with $I-K_s = 3.67$~mag in the Vega--based system, and
submillimeter flux densities of $S_{450} = 75\pm15$~mJy and $S_{850} =
17\pm 2$~mJy.     The intrinsic sub-mm flux density (corrected for
gravitational lensing magnification) is $S_{850} = 0.77$~mJy, a factor
of two fainter than the sensitivity of the deepest sub-mm field
surveys, and nearly an order of magnitude fainter than ``typical''
sub-mm--selected galaxies \citep{chap05}.   K04 provided the
spectroscopic redshift $z=2.5165\pm 0.0015$ based on the \ha\ emission
line in a near-IR spectrum and $z=2.514\pm 0.001$ from the ISM
absorption lines in an optical spectrum (rest--frame UV).
Furthermore, \target\ appears to be part of a larger  group at $z\sim
2.5$, with at least two other galaxies lensed by Abell~2218 with
$z=2.514$ based on their UV absorption lines.  

\citet[hereafter R08]{rig08} observed \target\ with \spitzer\ as part
of the time allocated to the guaranteed time observers (GTO).  The GTO
observations included imaging of Abell~2218 at 3.6--8~\micron\ from the
Infrared Array Camera  \citep[IRAC,][]{faz04}, and imaging at
24--160~\micron\ using the Multiband Imaging Photometry for \spitzer\
\citep[MIPS,][]{rie04}.   In the Appendix, we use the
available photometry from K04 and R08 to constrain the nature of the
stellar populations and to estimate the stellar mass of \target. 
R08 targeted component ``B''   of \target\
for mid--IR spectroscopy with \spitzer/IRS \citep{hou04} because this
source has the brightest IR flux densities, with $S_{24} =
1.16\pm0.10$~mJy.  Their IRS spectroscopy used the LL1 module, covering 19--38~\micron\ with
$R\sim 100$ resolution.   R08 fit template
IR SEDs from \citet{dal02} and \citet{cha01} to the 24, 450, and
850~\micron\ data (including a MIPS 70~\micron\ flux--density limit) to
derive a total IR luminosity integrated from 8--1000~\micron,
$L_\mathrm{IR} \equiv L(8-1000\micron)$.  They found $\lir  =
(7.6\pm1.9)\times 10^{11}$~\lsol, corrected for the
gravitational--lensing magnification, where the range reflects
differences in the template SEDs.  This IR luminosity corresponds to a
SFR, $\psi = 140\pm30$~\msol\ yr$^{-1}$,  using the relation in \citet{ken98}
for a Salpeter--like initial mass function (IMF) with lower and upper mass cutoffs of 0.1 and
100~\msol, respectively.   

\target\ appears to be undergoing an intense episode of star-formation
with no evidence for an active galactic nucleus (AGN).  K04 note that
the ratio [\ion{N}{2}]/\ha=$0.3\pm 0.1$ is consistent with ionization
from star--formation for stars of near solar metallicity
\citep[see][]{pet04}.  R08 show that the mid--IR spectrum contains
strong emission attributed to polycyclic aromatic hydrocarbon (PAH) molecules,
associated with dust heating from star--formation, with features
similar to local starburst galaxies.  The lack of any apparent silicon
absorption at 9.7~\micron\ in the spectrum argues against the presence
of an obscured AGN \citep[\eg,][]{spoon07}.  R08 analyzed archival
data from the  \textit{Chandra} X-ray Observatory, which provided
limits on the X-ray emission for this galaxy.  This flux upper limit
corresponds to a luminosity limit of $L(0.5-8\mathrm{keV}) < 5.0
\times 10^{42}$~erg~s$^{-1}$, corrected for the gravitational-lensing
magnification.  Compared to other sub-mm galaxies, the X-ray  upper
limit rules out the presence of an AGN unless it is very obscured
\citep[\eg][]{ale05}.

\section{OBSERVATIONS AND DATA REDUCTION}

\subsection{IRS SL2 Spectra}

For our \spitzer/IRS observations, we targeted component ``B'' of
\target, using the IRS SL2 module,
covering 5.2--8.7~\micron.  This is the same object targeted by R08, who
observed this galaxy with the IRS LL module.  Thus, the observations
here extend the spectral coverage of this target to the shorter
wavelengths, including the expected wavelength of \paa.

\begin{figure}[t]
\ifsubmode
\else
\epsscale{1.15}  
\fi
\vspace{5pt}
\plotone{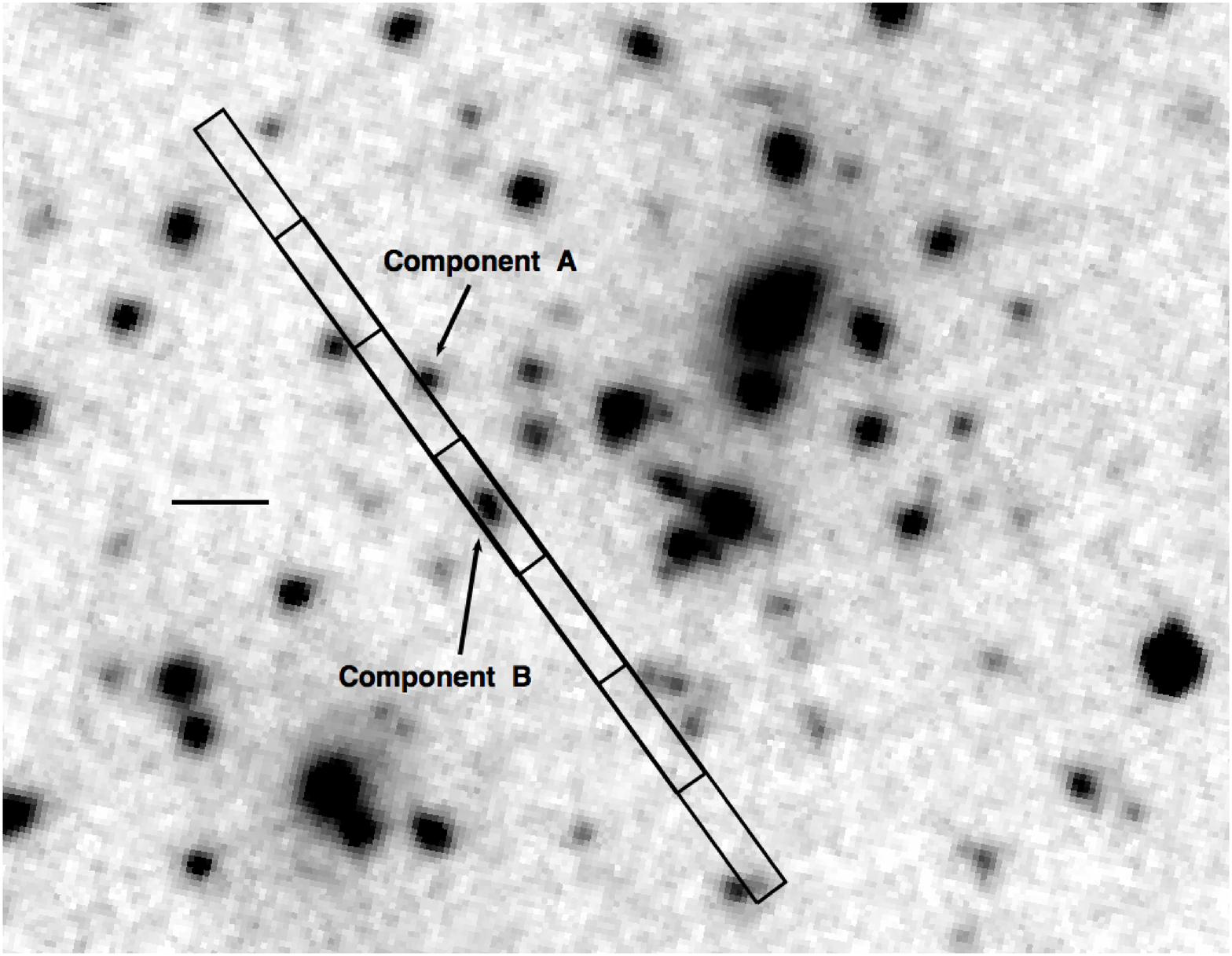}
\epsscale{1.0}  
\caption{ \spitzer\ IRS observations of \target.  The greyscale image shows the IRAC 5.8~\micron\ image of Abell 2218.   The four IRS SL2 slit positions and orientation for the observations of \target\ are indicated by the overlapping rectangles, centered on \target\ (component ``B'').  The horizontal black bar indicates a distance of 10 arcseconds.  Arrows indicate component ``B'' (the IRS spectroscopic target) and a component ``A'', a fainter counter image of this galaxy (using the notation of K04).\label{fig:sl2layout}}
\end{figure}

\begin{figure*}[th]
\ifsubmode
\else
\epsscale{1.0}  
\fi
\plotone{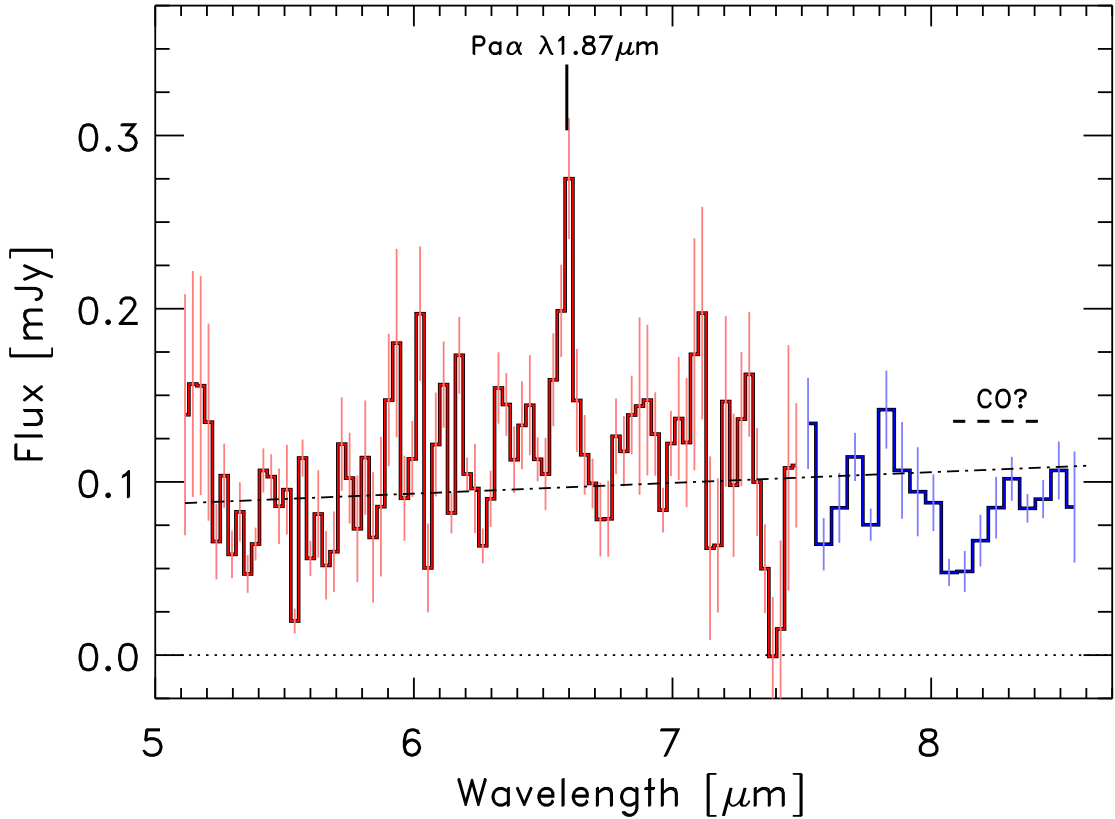}
\epsscale{1.0}  
\caption{ Extracted \spitzer\ IRS SL2 spectrum of \target, covering 5.0--8.4~\micron.  The
location of the \paa\ $\lambda1.8751$ emission line and possible CO absorption at a rest-frame of $z=2.515$ are indicated.  The dot--dashed line shows the continuum level \textit{predicted} from the IRAC 5.8 and 7.9~\micron\ flux densities, which is consistent with the measured continuum, excluding regions around the \paa\ emission line and excluding the possible CO absorption feature.  The red--shaded spectrum shows the SL2 2nd order spectrum.  The blue spectrum shows the SL2 1st order ``bonus'' spectrum.  The error bars correspond to the uncertainties measured from the IRS observations, propagated through the data reduction, spectral extraction, and combination procedures.\label{fig:1dspec}}
\end{figure*}

The observed flux density of \target\ at 5.8~\micron\ is $S_{5.8} =
92\pm9$~\ujy\ (R08).    For the redshift of our target, $z=2.515$, we
expect the \paa\ line to fall at 6.6~\micron.  We obtained our IRS
observations in Mapping mode, placing the object at four different
positions along the SL2 slit separated by 14\arcsec.
Figure~\ref{fig:sl2layout} shows the position of the four SL2 slit
positions on the IRAC 5.8~\micron\ image of the field for the epoch of
observations.   We observed \target\ using 60 cycles of 60~s
integrations, which we repeated at each of the four slit
positions. The total integration time is 4 hours, for which the
\spitzer\ Performance Estimation Tool predicts a signal--to--noise
ratio of $\simeq$6--7 per resolution element in the continuum.  In
practice, we find that systematics reduce this slightly (see below).

Observations were obtained on 2008 Feb 22-23 during a single AOR in
order to prevent alignment uncertainties from different campaigns
possibly at very different spacecraft orientations.  We used high
accuracy peak-up centroiding with the IRS blue filter to minimize any
deviations in the absolute telescope pointing.  We chose a peak--up
star from the 2MASS catalog, against which we had matched the
astrometric coordinates for \target.    

We reduced the data starting with the S17.2.0 \spitzer\ IRS pipeline
data, which produced basic calibrated data (BCD) files.  Our data
contained some very strong cosmic rays which adversely affected the
stray light correction in the IRS pipeline.  However, we determined
that the stray light effects are negligible in our data as the peak-up
arrays do not contain bright sources during our exposures. We
therefore used the BCD data products without the stray light
correction (the BCD files \texttt{f2ap.fits} and \texttt{f2unc.fits}).

The dominant background component at 5--8~\micron\ is Zodiacal light.
Our observations included very long integrations of faint sources
(these are currently the deepest SL2 observations taken of
\textit{any} object of which we are aware).  Therefore we include some
additional steps to remove the sky background and detector effects.
We generally follow the post--BCD reduction steps used by
\citet{tep07} and \citet{pop08} in their analysis of deep IRS/LL data.
We looked for indications of latent charge in the array
during our long exposures, as latent charge accumulation is observed
in long IRS/LL exposures \citep[e.g.,][]{tep07}.  However, we observe
no increase in the median number of ADU s$^{-1}$ over the course of a
60~min SL2 observation. Thus, there is no evidence for appreciable
latent charge in the SL2 array over our integrations.

We next identified and cleaned known bad and hot pixels using the SSC
task
IRSCLEAN\footnote{http://ssc.spitzer.caltech.edu/postbcd/irsclean.html}
with the known warm--pixel mask for our IRS campaign (IRSX008900).  We
also identified other ``rogue'' and ``warm'' pixels as pixels with
abnormally high variance.  We performed this latter task using an
automated routine which computed the variance of each pixel over all
the BCDs of an observation.  We flagged rogue pixels as those pixels
with high variance ($> 4 \times \sigma^2$).  We then used IRSCLEAN to
interpolate over these pixels. 

We constructed a sky spectrum for \target\ at each of the four slit
positions by combining the BCDs of the three other slit positions.  To
create the sky image, we took the median of the stack of each pixel
after rejecting outliers using a sigma--clipping algorithm.  We
performed this process iteratively, masking out the location of
\target\ during subsequent iterations (\target\ is the only source we
identify in the 2D spectrum).   We then subtracted the sky frame from
each BCD and coadded the BCDs at each slit position.  As a last step,
we reran IRSCLEAN on the combined images for each slit position to
clean any remaining hot pixels.   These steps produced four 2D
spectroscopic images for \target, one at each of the four slit
positions. 

We extracted 1D spectra at each slit position using the \spitzer/IRS
custom extraction (SPICE)
software\footnote{http://ssc.spitzer.caltech.edu/postbcd/spice.html}
with an optimal extraction window.   We extracted spectra from both
the 2nd spectral order (covering 5.2--7.7~\micron) and the ``bonus''
1st spectral order (covering 7.3--8.7~\micron).   SPICE provides the
extracted 1D science spectrum and the propagated errors from the IRS
observation.  For each science spectrum, we also extracted a sky
spectrum using the same SPICE parameters, offset from \target.  These
sky spectra provide consistency estimates for the error derived from
the science spectra.  We found these to be consistent, and we adopted
the latter here in subsequent analysis.  We then combined the four 1D
spectra as a weighted mean, using the weights derived from the error
spectra.  The estimate of the variance on the combined 1D science
spectrum is the inverse sum of the weights.  Figure~\ref{fig:1dspec}
shows the combined 1D spectrum.

\target\ is marginally resolved in the IRAC 5.8 and 8.0~\micron\
images (channels 3 and 4).   We estimated the angular resolution using
three stars in the field identified from \hst/ACS imaging.   For these
three stars we measured a mean FWHM=$1\farcs8$ in the channel 3 image.
In comparison, for \target\ we measure a FWHM=$2\farcs0$ in this
image.  Most of the elongation of \target\ appears to be aligned with
the orientation of the SL2 slit for our observations, as illustrated
in figure~\ref{fig:sl2layout}.  Although \target\ appears resolved,
for the spectral extraction and flux calibration we nevertheless
assumed \target\ to be a point source.  This is reasonable because IRS
SL2 has large pixels (1.8\arcsec\ pixels) compared to IRAC
(1.2\arcsec\ pixels), so the spatial resolution is substantially worse
compared to IRAC, and \target\ is at best marginally resolved.  In
principle, if it is resolved spatially, this could affect the flux
calibration of the extracted spectrum. However, we measure a continuum
of \target\ from our SL2 spectrum that is within 1\% of that expected
from the measured IRAC 5.8 and 7.9~\micron\ flux densities (see
figure~\ref{fig:1dspec}).     Because pointing--induced slit
throughput variations may produce uncertainties in the absolute flux
measurements of 10--15\%, we consider this agreement fortuitous.
Nevertheless,  we conclude that our extracted spectrum corresponds to
the spatially integrated emission of \target, and we make no
corrections because of the extended nature of this source.
Furthermore, because the expected flux densities and measured
continuum are essentially equal, we also make no additional
corrections to the measured flux for light falling outside the SL2
slit.   

\subsection{MIPS 70~\micron\ Imaging}\label{section:data_mips}

\target\ has \spitzer/MIPS 70 and 160~\micron\ data from the GTO
observations.  Using these, R08 placed an upper limit on the
70~\micron\ emission of $S_{70} < 7$~mJy.   Since then, deeper MIPS
70~\micron\ data have become available as part of \spitzer\ program ID
30823.  These new observations reach an exposure time of 5245~s at the
deepest point, substantially longer than the GTO observations.  We
obtained the raw MIPS 70~\micron\ images from the \spitzer\ archive
and reduced them using the GTO Data Analysis Tool  \citep{gor05}
following the steps described in \citet{dole04}.  We have applied
further steps to mask sources detected in the image during the
background subtraction steps, which greatly improves the final image
quality.  The final image achieves a limiting flux density of
$\sigma_{70}$=0.5~mJy for point sources.

We detected point sources in the image using a weighted detection map
and performed point--response function (PRF) weighted
photometry. Because there are not enough bright point sources in the
image to construct an empirical PRF from the image, we used the
empirical PRF constructed from sources in wider area (0.5~deg$^2$)
extragalactic fields \citep{pap07}.  Photometry is performed
simultaneously on crowded sources using a version of the DAOPHOT
software \citep{ste87}, following the procedure from \citet{pap07}. 

\begin{figure}[t]
\ifsubmode
\else
\epsscale{1.23}  
\fi
\plotone{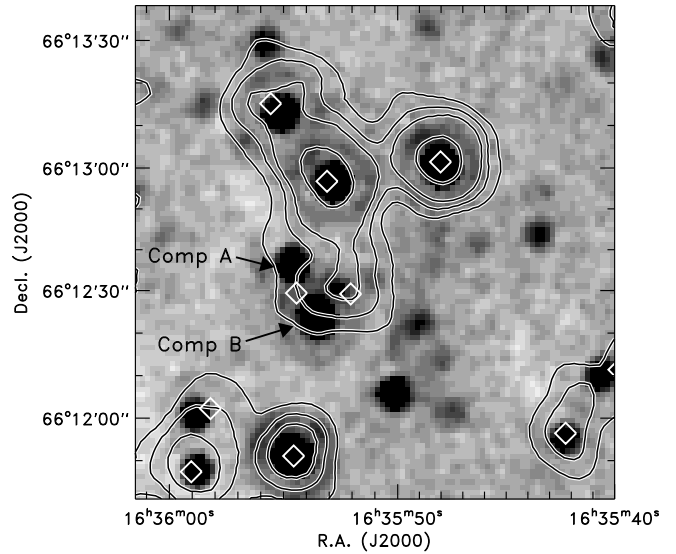}
\epsscale{1.0}  
\caption{ \spitzer/MIPS 24~\micron\ image of Abell 2218.    \target\ (Comp B, in the notation of K04) and its counter image (Comp A) are labeled.  Contours show the \spitzer/MIPS 70~\micron\ flux density, at levels of 1, 2, 3, 6$\times$ the rms noise.   Diamonds denote sources detected in the 70~\micron\ data.   The 70~\micron\ source most closely associated with \target\ is blended with its counter image and partially with the other nearby source.\label{fig:70umcontour}}
\end{figure}

\target\ is marginally confused in the 70~\micron\ image.
Figure~\ref{fig:70umcontour} shows the contours of the 70~\micron\
image overlaid on the 24~\micron\ image.  Given the substantially
poorer resolution at 70~\micron, the detected source corresponding to
\target\ is likely blended with adjacent sources.  We measured a flux
density of $S_{70} = 2.56\pm 0.90$ for the source most closely
associated with \target\ (uncorrected for gravitational
magnification).   At the 70~\micron\ angular resolution of \spitzer,
\target\ is blended with a counter image of itself (component ``A'' in
the notation of K04), and possibly another source.  Given that the
ratio of the 24~\micron\ flux densities of these components is
$S_{24}(\mathrm{B})/S_{24}(\mathrm{A})$=1.6, we expect a similar ratio
at 70~\micron\ because presumably component A and B have the same
24~\micron--to--70~\micron\ flux ratios (although K04 note that
component B  has a redder $I-K$ color, and thus could have a higher
$S_{70}/S_{24}$).    To improve our flux--density measurements, we
simultaneously measured point--source photometry in the 70~\micron\
image using \textit{a priori} positions for sources detected in the
24~\micron\ image.   However, this method was very sensitive to
uncertainties in the absolute astrometry between the 24 and
70~\micron\ images, and yielded a flux density for component B
ranging from $S_{70} = 0.0$~mJy to 1.5~mJy ($\pm 1.0$~mJy).   Therefore
in the analysis below, we use the direct photometry, and ascribe all
the 70~\micron\ emission to \target\ (component B).  We note that the
true 70~\micron\ flux density may be lower by as much as 40\% assuming
the $S_{70}/S_{24}$ ratio above.  As we will discuss below, the
measured 70~\micron\ flux density implies a rest--frame
$L(24\micron)$--to--$L(\paa)$ ratio lower than that for local
IR--luminous galaxies with comparable bolometric luminosities.
Therefore, our assumption makes this result conservative.

% Note to self:  From IMEXAM, the flux ratio of 24 um of component B
% to component A is 217. / 139.3 = 1.58.   Given that S70=2.56 mY,   % then arguably comp B has S70=1.6 and comp A has S70=1.0 mJy. 

\section{ANALYSIS}\label{section:analysis}

\subsection{Paschen--$\alpha$ Emission}

\ifsubmode
\begin{deluxetable}{cccccccccc}
\rotate
\else
\begin{deluxetable*}{cccccccccc}
\fi
\tablewidth{0pt}
\tablecaption{Derived Quantities \label{table}}
\tablehead{\colhead{$\lambda_C$} & \colhead{} &
\colhead{EW$_0$}  & \colhead{$F(\paa)$} & \colhead{$L(\paa)$}
& \colhead{$A(V)$} & \colhead{$A(\paa)$} & \colhead{$L(\paa)_\mathrm{cor}$} & \colhead{SFR$$} \\
\colhead{(\micron)} & \colhead{$z$} &
\colhead{(\AA)}  & \colhead{($  10^{-16}$ erg s$^{-1}$ cm$^{-2}$)} & \colhead{($  10^{42}$ erg s$^{-1}$)}
& \colhead{(mag)} & \colhead{(mag)} & \colhead{($  10^{42}$ erg s$^{-1}$)} &
\colhead{(\msol\ yr$^{-1}$)} \\
\colhead{(1)} & \colhead{(2)} & \colhead{(3)} &
\colhead{(4)}  & \colhead{(5)} & \colhead{(6)}
& \colhead{(7)} & \colhead{(8)} & \colhead{(9)} }
\startdata
6.591 $\pm$ 0.006  & 2.515 $\pm$ 0.003 &  
363 $\pm$ 56 & $8.5 \pm 1.4$  & $2.05 \pm 0.33$ & 3.6 $\pm$ 0.4 & 
0.27 $\pm$ 0.03 & $2.57 \pm 0.43$ & $171 \pm 28$
\vspace{3pt}\enddata
\tablecomments{ (1) Centroid wavelength of \paa\ line, (2) measured
redshift, (3)
rest--frame equivalent width of the \paa\ line, (4) measured line flux, with no correction for the
gravitational lensing magnification or dust extinction, (5) line
luminosity, corrected for the gravitational lensing magnification $\mu=22$
(K04), (6) nebular extinction from \ha\ and \paa\
measurements using the \citet{cal00} law, (7) corresponding extinction at 1.8751~\micron,  (8)  line luminosity, corrected for the gravitational
lensing magnification and dust extinction, (9) total SFR, corrected
for the gravitational lensing magnification and dust extinction. }
\ifsubmode
\end{deluxetable}
\else
\end{deluxetable*}
\fi

Figure~\ref{fig:1dspec} shows the combined IRS/SL2 spectrum.   The
dot--dashed line in the spectrum shows the predicted continuum from
the measured IRAC photometry at 5.8 and 8.0~\micron,
$S_{5.8}=92$~\ujy\ and $S_{7.9}=110$~\ujy, which agrees with the
measured continuum to within 1\%.   The median signal-to-noise ratio
of the continuum is S/N$\approx$5 from 5.5--7~\micron.  

We identify the emission line at 6.591~\micron\ as \paa\ at $z=2.515$,
consistent with the expected redshift derived by K04.   This is the
highest redshift detection of \paa\ in any galaxy of which we are
aware \citep[\cf,][]{sia08,sia09}.  To measure the line parameters, we
simultaneously fit the continuum and the line, where we model the
latter as a Gaussian.   The Gaussian fit has a measured full-width at
half maximum, FWHM=$0.067\pm0.014$~\micron.  This is comparable to
the IRS/SL2 resolution, therefore we conclude the line is unresolved.   To derive
the significance of this detection we generated a series ($10^3$) of
simulated spectra using the data and a random value taken from a
Gaussian distribution with $\sigma$ equal to the derived uncertainty
on each data point.   We then remeasure the line flux in each
simulated spectrum, and take the inner 68\% of the simulated
distribution for each parameter as the uncertainty.  The measured
quantities are given in Table~\ref{table}.     We derive a redshift
for the \paa\ line of $z=2.515\pm0.003$, which agrees within the
measured uncertainties of K04 and R08 using spectral features in the
optical, near--IR, and mid--IR wavelength ranges.

We measure a \paa\ line flux of $(8.6\pm1.2)\times10^{-16}$ erg
s$^{-1}$ cm$^{-2}$, uncorrected for the gravitational lensing
magnification or dust attenuation.   We estimate the amount of dust
attenuation affecting the nebular gas by comparing the ratio of the
\paa\ line flux to the \ha\ line flux provided by K04 using $K$--band
spectroscopy.     Note that the K04 measurement of \ha\ has not been
corrected for extended emission beyond their spectroscopic slit of width 0\farcs76.  We
estimate that the slit width used in the $K$--band spectroscopy of K04
would miss up to 35\% of the light from \target.  If we applied this
(maximal) correction to the \ha\ flux, then it would decrease the
extinction $A(\paa)$ by 0.03~mag, and decrease the
extinction--corrected \paa\ luminosity by $\approx$2\%.  Because these
corrections are small, we do not apply any correction for the light
falling outside the slit of the $K$--band spectroscopy.   Using the
\citet{cal00} attenuation law, we estimate the extinction to be $A(V)
= 3.6\pm 0.4$~mag, which corresponds to $A(\paa)=0.27\pm 0.03$~mag.
Other attenuation laws that we tested \citep{rie85,car89,dra89,dop05}
correspond generally to higher extinction estimates by as much as $\delta A(V) \approx
0.5$~mag compared to that of \citet{cal00}.   However, \citet{cal05}
showed that compared with other attenuation laws their attenuation law reproduces better the UV
colors of starforming \ion{H}{2} regions with similar extinction and
properties as that derived here for \target.  Therefore, we adopt the Calzetti et al.\ law for our
analysis here.  As we will discuss below, the ratio of the
mid--IR luminosity to the extinction--corrected $L(\paa)_\mathrm{cor}$ is
lower relative to local IR galaxies of comparable luminosity.   Using
other attenuation laws would increase the \paa\ line luminosity by
$\approx$30\%.  Therefore using the lower extinction value provided by
the Calzetti law also provides a conservative choice.    Nevertheless,
the dust--extinction correction remains a systematic
uncertainty.\footnote{We have learned that a recent reanalysis of the
K04 near-IR spectroscopy (Richard et al., in preparation) provides a
\ha\ flux that is $\approx$4$\times$ fainter than that of K04.  If
this flux is correct, it would increase the extinction to
$A(\ha)=4.9$~mag and $A(\paa)=0.45$~mag, and would increase the
extinction--corrected \paa\ luminosity (and SFR) by $\approx 18$\%.
However, none of the main conclusions here would be affected
significantly.}

We derive a \paa\ line luminosity of  $L(\paa)_\mathrm{cor}
  = (2.57\pm 0.43) \times 10^{42}$~erg s$^{-1}$  after applying the extinction
correction, $10^{0.4 A(\mathrm{Pa}\alpha)} = 1.28$, the gravitational
lensing magnification, $\mu=22$, and the luminosity distance, $d_L =
2.054\times10^4$~Mpc, at $z=2.515$ for the default cosmology.   This
luminosity corresponds to an ionizing continuum flux of
$Q(H^0)=(1.6\pm 0.3) \times 10^{55}$~$\gamma$ s$^{-1}$ \citep{ost89,ken98}.  For
an IMF with a Salpeter--like slope from 0.1 to 100~\msol, the implied
SFR is $171\pm 28$~\msol\ yr$^{-1}$.  The statistical uncertainty on
the SFR is 17\%, which is the highest accuracy on a SFR derived for a
high--redshift galaxy to date.     Note that while this uncertainty neglects the
uncertainty on the gravitational lensing magnification ($\approx
10$\%, see \S~2), the lensing is expected to be achromatic, and it will systematically scale all derived luminosities and SFRs.

\subsection{ CO Absorption}

The IRS/SL2 1st--order ``bonus'' spectrum shows a possible absorption
feature at 8--8.5~\micron\ (see figure~\ref{fig:1dspec}).  At
$z=2.515$ this corresponds to $\approx$2.3--2.4~\micron, and would
imply strong molecular CO absorption.  If this feature is real, it
would be the first time it has been observed in a galaxy with any
significant redshift.  CO absorption occurs in the atmospheres of red
supergiant stars (K-- and M--type), primarily from post--main-sequence
O--type stars (e.g., Doyon, Joseph, \& Wright 1994; Ridgway,
Wynn-Williams, \& Becklin 1994; Goldader \etal\ 1995).  Given the
quality of our SL2 spectrum, a detailed analysis of the CO absorption
is cautionary.  Nevertheless, we derive a spectroscopic CO index of
CO$_\mathrm{sp} = 0.29\pm0.05$ using the definition of \citet[see
also Smith et al.\ 1996]{doy94}.  Combined with the slope of the
rest--frame near--IR continuum, this places \target\ in the ``[Dust--]
Reddened Starburst Population'' locus of \citet{rid94}, implying
that the starburst in this galaxy dominates the rest--frame near--IR
emission.  

If the strength of the CO absorption feature in \target\ is real, then
the age of the starburst corresponds to the time it takes O--type
stars to enter the red supergiant phase.  The fact that the \ion{H}{1}
recombination lines remain strong requires ongoing early--type
(primarily O--type) star formation.  The combination of these facts
implies an age for the starburst of $\sim$10--50~Myr \citep{doy94}.

\subsection{The Mid--Infrared Luminosity}

Using the MIPS 24 and 70~\micron, and IRS/LL observations, we derive
monochromatic luminosities at rest--frame 8 and 24~\micron,
$L(8\micron) \equiv \nu L_\nu(8\micron)$ and $L(24\micron) \equiv \nu
L_\nu(24\micron)$.  Below we compare our results for \target\ against
datasets of local galaxies where the above quantities are derived
using photometry measured from IRAC 8~\micron\ and MIPS 24~\micron.
Therefore, we compute $L(8\micron)$ and $L(24\micron)$ which match the
rest--frame IRAC and MIPS bandpasses as closely as possible.

\begin{figure}[b]
\ifsubmode
\else
\epsscale{1.2}
\fi
\plotone{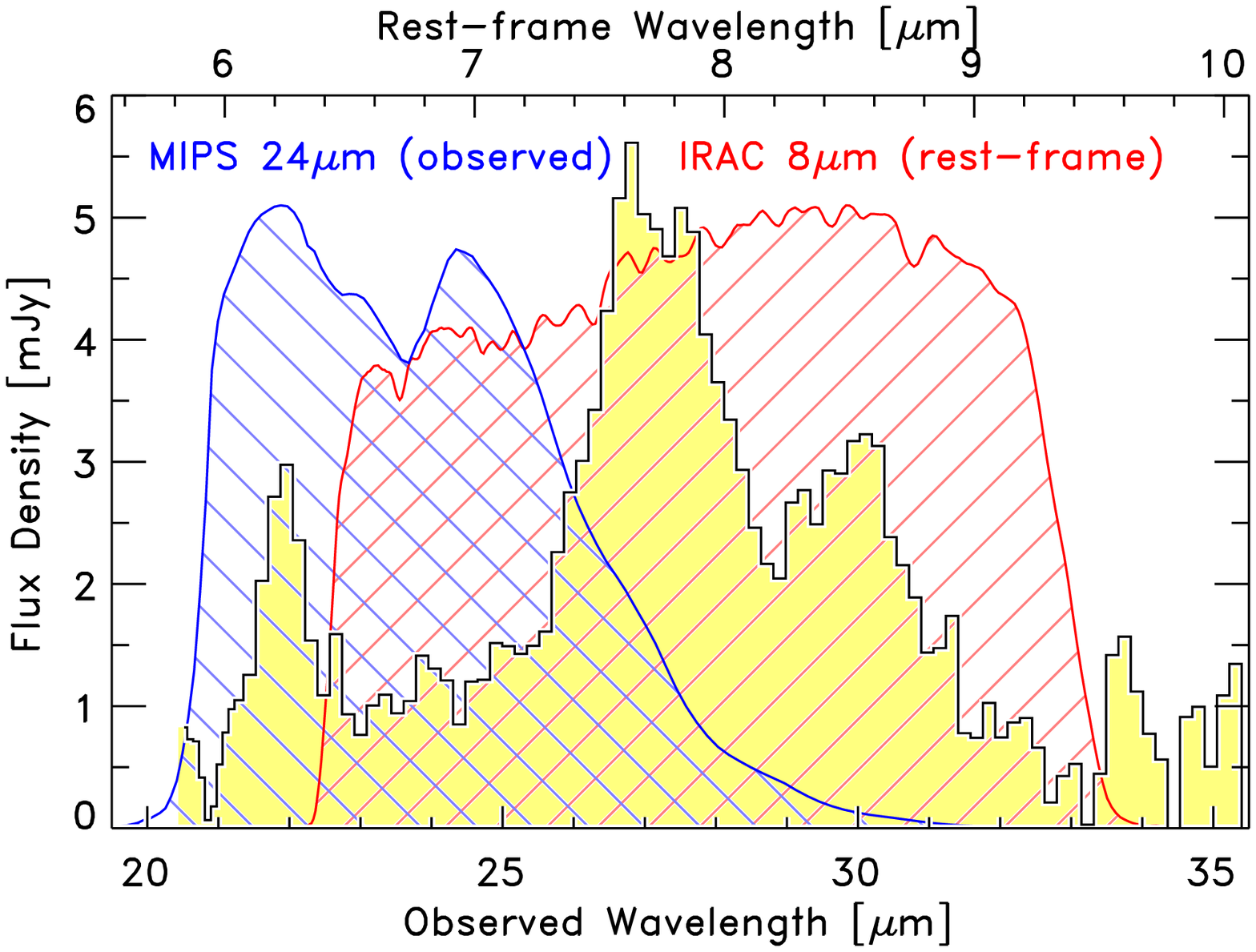}
\caption{IRS/LL spectrum from 20--35~\micron\ for \target\ from R08.
The IRS spectrum is shown as the yellow--filled histogram (R08).  The
blue--hashed region denotes the MIPS 24~\micron\ bandpass in the
observed frame at $z=2.515$.  The red--hashed region denotes the IRAC
8~\micron\ bandpass in the rest-frame.  We integrated the IRS/LL
spectrum with the 8~\micron\ bandpass in the rest-frame to derive the
rest--frame 8~\micron\ flux density. This differs significantly from
the observed MIPS 24~\micron\ flux density because of the contribution from strong spectral features in this wavelength region attributed to PAHs, especially at 6.2, 7.7 and 8.6~\micron.\label{fig:irsll}}
\end{figure}

At $z=2.515$ the  MIPS 24~\micron\ photometry corresponds to
rest--frame 6.8~\micron.  To
convert this to rest--frame IRAC 8~\micron\ we use the IRS/LL spectrum
of \target\ from 20--38~\micron\ from R08, corresponding to
rest--frame 5.7--10.8~\micron.    Figure~\ref{fig:irsll} shows the
IRS/LL spectrum of \target\ from R08 overlaid with the MIPS
24~\micron\ bandpass in the observed frame and the IRAC 8~\micron\
bandpass in the rest frame.  We integrate the IRAC 8~\micron\
transmission function with the IRS/LL spectrum (shifted to the
rest--frame) to derive a rest--frame 8~\micron\ flux density,
$f_{\nu,0}(8\micron)$=$1.5\pm0.2$~mJy, where the subscript ``0'' denotes the
rest--frame quantity.  This corresponds to
$f_{\nu,0}(8\micron)$=$70\pm 10$~\ujy\ corrected for the gravitational
magnification.  In contrast, R08 measured a 24~\micron\ flux density
of $S_{24}=1.16$~mJy.  This difference is primarily a consequence of
the fact that at $z=2.515$ the MIPS 24~\micron\ bandpass is mostly
insensitive to the strong 7.7~\micron\ and 8.6~\micron\ PAH features,
which dominate the mid-IR spectrum of \target.  These PAH features are
included in the rest--frame IRAC 8~\micron\ bandpass, which results in
the much higher flux density when averaged over this bandpass.

\begin{figure}[t]
\ifsubmode
\epsscale{1.0}  
\else
\epsscale{1.2}
\fi
\plotone{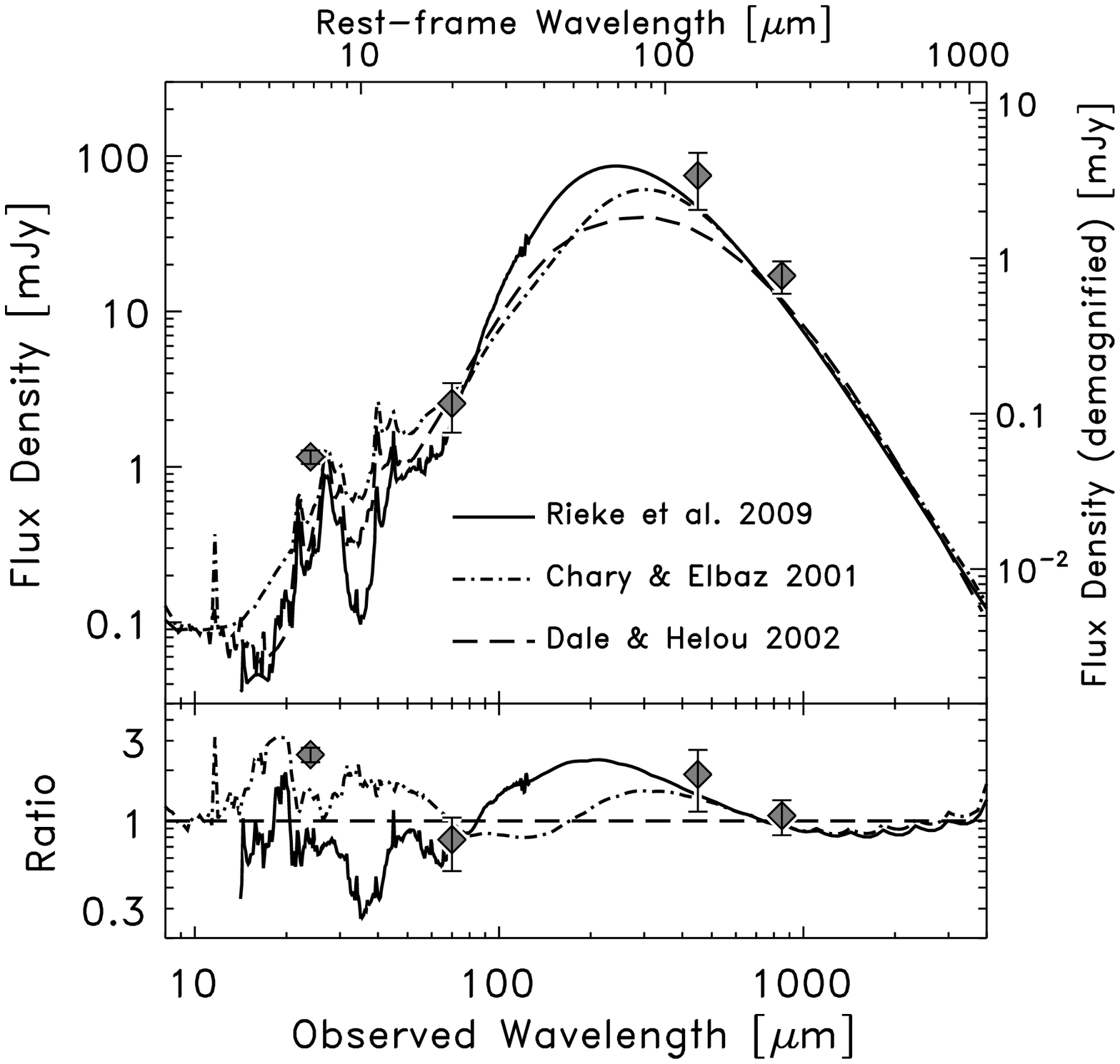}
\epsscale{1.0}  
\caption{ Infrared SED of \target.  In the top panel, the data points show the MIPS 24 and 70~\micron, and SCUBA 450 and 850~\micron\  measured flux densities.  The right axis shows the flux densities after correcting for the gravitational lensing magnification.  The top axis shows the rest--frame wavelength for $z=2.515$.  The curves show IR SED template fits to the 70, 450, and 850~\micron\ flux densities, using templates from \citet[solid black line]{rie09}, \citet[dashed line]{dal02}, and \citet[dot-dashed line]{cha01}.   Formally the Chary \& Elbaz and Rieke et al.\ templates provide better fits to the data.  However, uncertainty in the total IR luminosity on the order of a factor of two remains owing to differences in the templates.   The range of the implied total IR luminosity, $\lir \equiv L(8-1000\micron)$, ranges from $5-10 \times 10^{11}$~\lsol, depending on the model adopted.   To improve these constraints requires flux density measurements  at observed wavelengths of $\approx$150--250~\micron, which will constrain the peak of the thermal dust emission.  The bottom panel shows the ratio of the best-fit models and data points to the best-fit model of Dale \& Helou.  Note that all of the best--fit models to the far--IR data underpredict the observed 24~\micron\ flux density by factors of $>$2. \label{fig:irsed}}
\end{figure}

We measured the total IR luminosity for \target\ by fitting a suite of template IR SEDs
\citep{cha01,dal02,rie09} to the  70, 450, and 850~\micron\
photometry, as illustrated in figure~\ref{fig:irsed} (see also \S~5.1).   The best--fit templates correspond 
to a total IR luminosity, $L_\mathrm{IR} = (5-10) \times 10^{11}$~\lsol, corrected for the gravitational lensing magnification.  We estimated the rest--frame 24~\micron\ flux density for \target\ by integrating these best--fit IR SEDs with the MIPS 24~\micron\ transmission function.  This yielded, $f_{\nu,0}(24\micron) = 0.24\pm
0.09$~mJy, corrected for the gravitational lensing magnification.
There is an additional small systematic uncertainty resulting from
differences in the template IR SEDs, corresponding to
$\sigma_\mathrm{sys}=0.03$~mJy.    The error bar here is dominated by
the uncertainty on the 70~\micron\ photometry, which is the closest
band to the rest--frame 24~\micron\ datum.   Furthermore, we remind
the reader that the 70~\micron\ flux density may be lower by as much
as 40\% (see \S~\ref{section:data_mips}), which would decrease the
24~\micron\ luminosity.

\begin{figure*}[t]
\epsscale{1.15}  
\plottwo{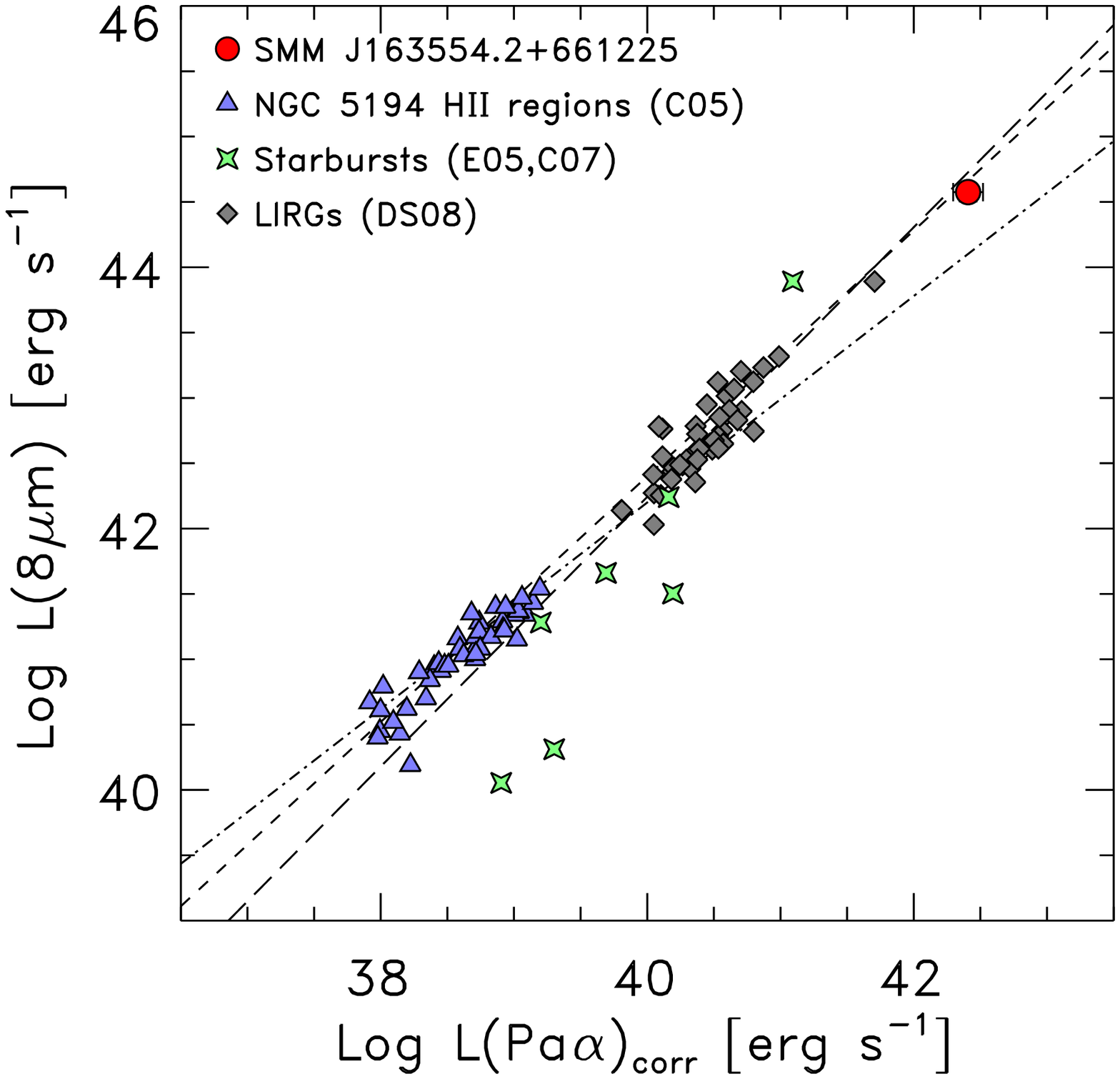}{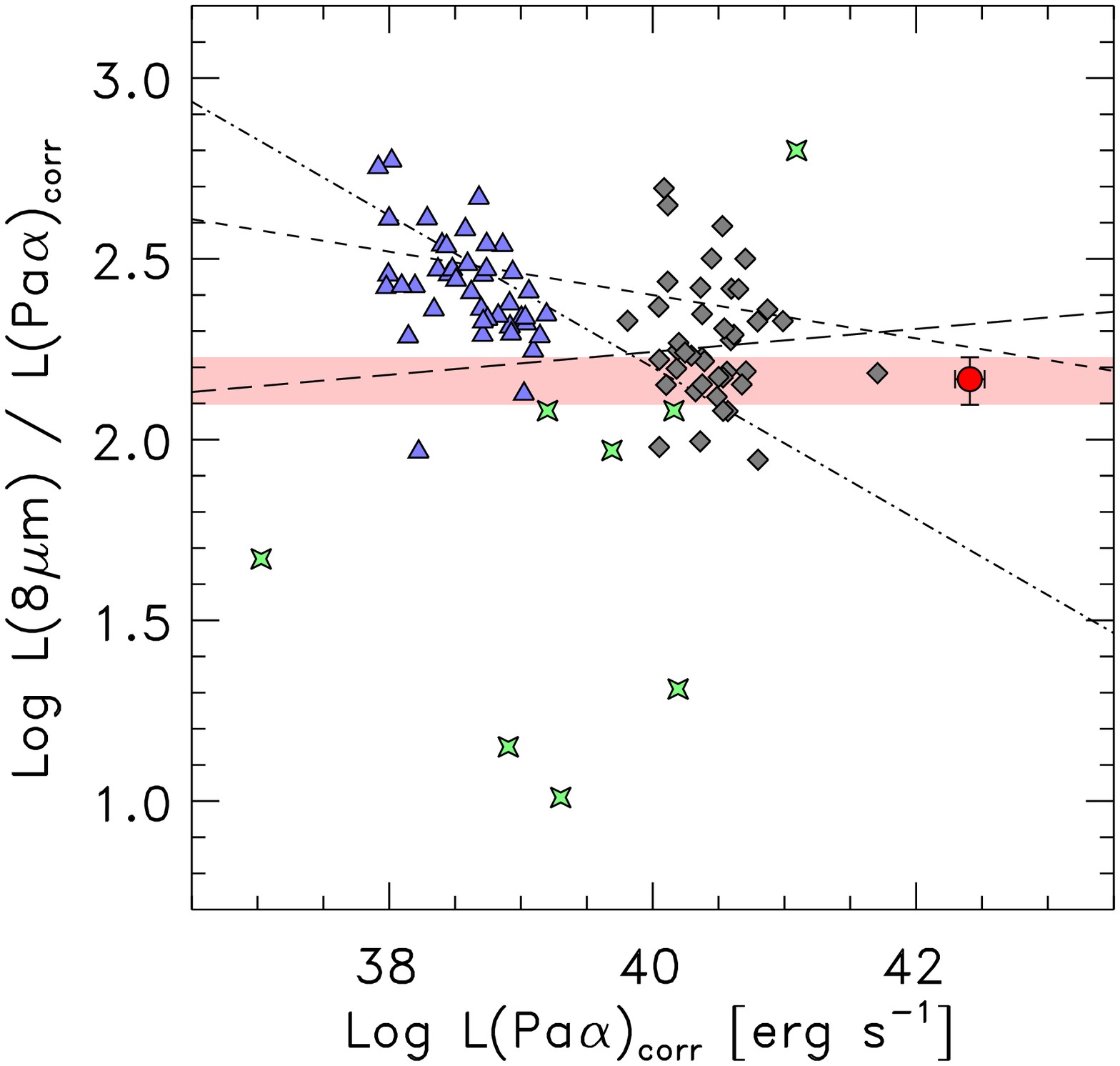}
\epsscale{1.0}  
\caption{ \paa\ luminosity versus the monochromatic luminosity at
rest--frame 8~\micron.   The left panel shows $L(\paa)$ versus
$L(8\micron)$ and the right panel shows $L(\paa)$ versus the ratio
$L(8\micron)/L(\paa)$.  In both panels the symbols and lines are the
same.   The large, red--filled circle shows the  measured value for
\target.  The light-red--shaded area indicates the error bar on the
ratio for \target.   Diamonds show  local luminous IR galaxies ($\lir
= 10^{11} - 10^{12}$~\lsol) from the sample of \citet[DS08]{dia08}.
Triangles show individual star--forming regions in M51
\citet[C05]{cal05}. Stars show starburst and low--metallicity galaxies
from \citet[C07]{cal07} and \citet[E05]{eng05}.     The dot--dashed
line shows the best-fit relation to individual star--forming H~II
regions in M51 from \citet{cal05}.  The short--dashed line shows the
best--fit relation derived for star--forming galaxies and individual
H~II regions from \citet{cal07}, which is similar the relation for luminous IR galaxies derived by
\citet[long--dashed line]{dia08}.  The 8~\micron\ and \paa\ luminosities for \target\ are consistent with the extrapolated relationship for local star--forming regions and star--forming galaxies.  \label{fig:8um}}
\end{figure*}

We convert the rest--frame flux densities to luminosities using,
$L(\lambda) = \nu f_{\nu,0}(\lambda) \times (4 \pi d_L^2) \times \mu^{-1}$, where $\mu$ is the gravitational lensing magnification, and 
$d_L$ is the luminosity distance.    Applying this formula to
the 8 and 24~\micron\ rest-frame quantities derived above yields
$L(8\micron) = (3.8\pm 0.6) \times 10^{44}$~erg s$^{-1}$ and
$L(24\micron) = (4.3 \pm 1.6)\times 10^{44}$~erg s$^{-1}$.

\section{DISCUSSION}\label{section:discussion}

\begin{figure*}[t]
\epsscale{1.15}  
\plottwo{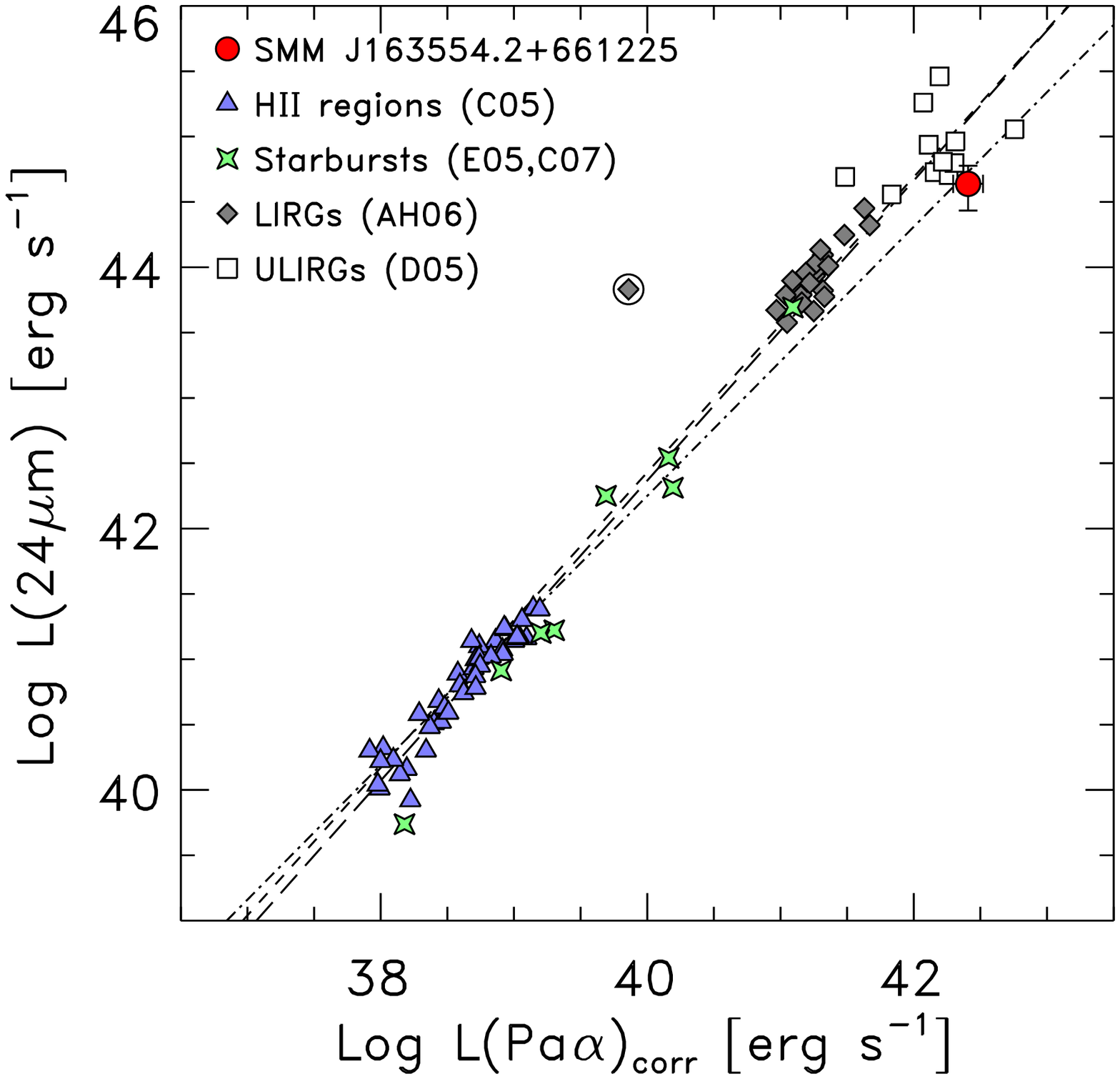}{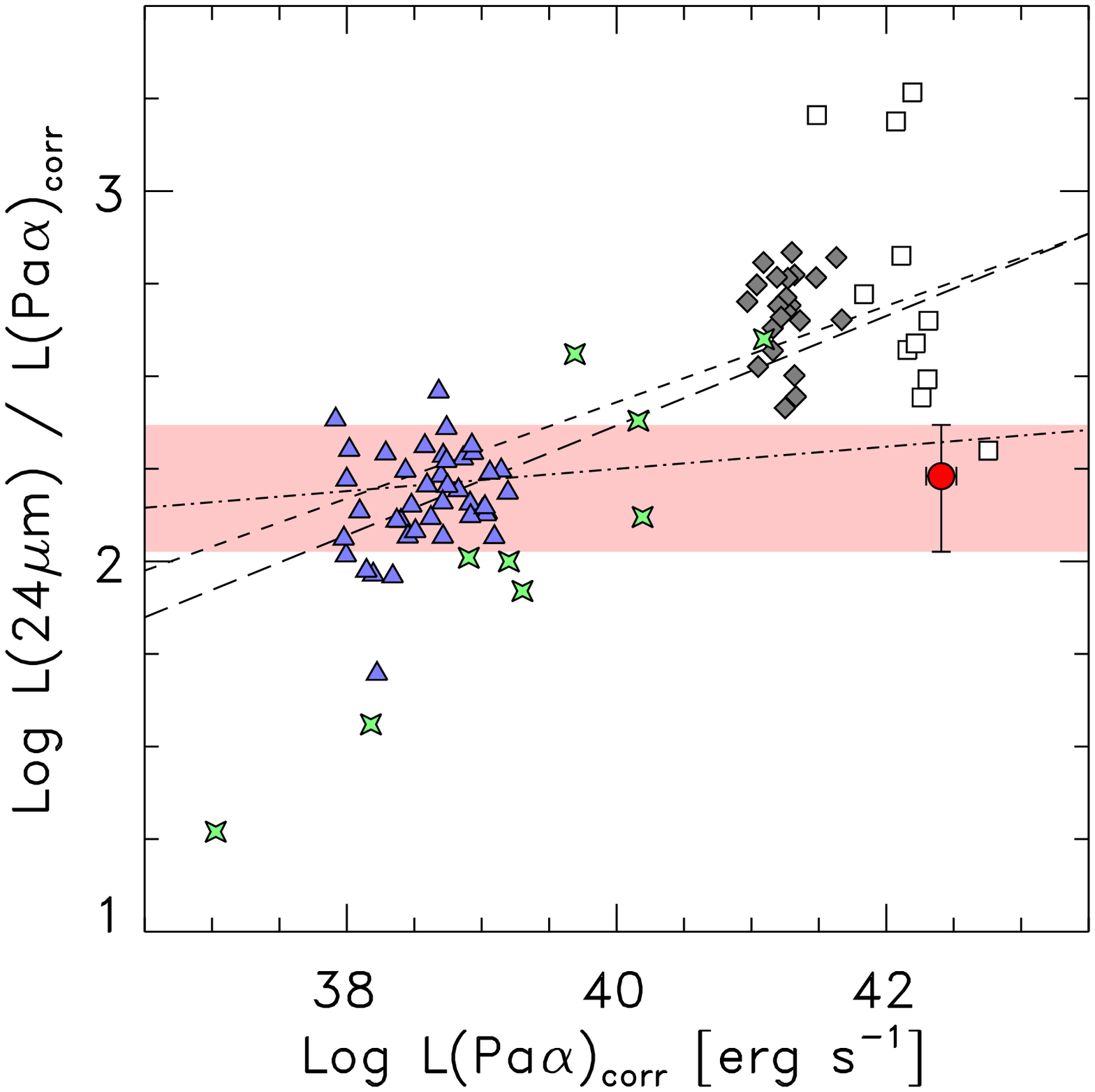}
\epsscale{1.0}  
\caption{ \paa\ luminosity versus the monochromatic luminosity at
rest--frame 24~\micron.  The left panel shows $L(\paa)$ versus
$L(24\micron)$ and the right panel shows $L(\paa)$ versus the ratio
$L(24\micron)/L(\paa)$.  In both panels the symbols and lines are the
same.   The measured value for \target\ is shown as the large, red--filled
circle.  The light-red--shaded area indicates the error bar on the
ratio for \target.  Note that $L(24\micron)$ for \target\ may be lower
by $\approx$40\% owing to crowded photometry (see \S~3.2).    Squares
show local ultraluminous IR galaxies (ULIRGs, $\lir \ge 10^{12}$~\lsol) from
\citet[D05]{dan05}, with $L(24\micron)$ estimated using the method
discussed in the text.  Diamonds show local luminous IR galaxies
(LIRGs, $\lir = 10^{11} - 10^{12}$~\lsol) from the sample of
\citet[AH06]{alo06}; the diamond indicated by the circle corresponds
to the galaxy IC~860, which is problematic according to A06.  The
range of the ordinate in the right panel excludes IC~860.  Triangles
show local star--forming H~II regions in M51 \citet[C05]{cal05}.
Stars show starburst and low--metallicity galaxies from
\citet[C07]{cal07} and \citet[E05]{eng05}.     The short--dashed line
shows the best fit relationship to local luminous IR galaxies from
\citet{alo06}.  The long--dashed line shows the best-fit to local
galaxies and star--forming regions from \citet{cal07}.   The
short--dashed line shows the best-fit relation for individual H~II
regions \citet{cal05}.  The rest--frame 24~\micron\ luminosity for
\target\ is significantly lower (by $\approx 0.3-0.5$~dex) with
respect to local galaxies of similar \paa\ luminosity.  In contrast,
the $L(24\micron)/L(\paa)$ for target is consistent with that found in
individual star--forming regions.   \label{fig:24um}}
\end{figure*}

The observations for \target\ provide independent estimates for the
total SFR.     In particular, we compare the \paa\ luminosity which
stems from the ionized gas in \ion{H}{2} regions and traces the number
of ionizing photons, to estimates from the mid--IR and total IR
luminosities, which measure primarily the dust--reprocessed emission
from massive stars \citep[\eg,][]{ken98,kew02}.  As discussed in
\S~4.1, the \paa\ line luminosity corresponds to a SFR $\psi = 171\pm
28$~\msol\ yr$^{-1}$.    

\subsection{The \paa\ Luminosity Compared to the Total IR Luminosity}

Local star--forming galaxies and star--forming regions show a tight
correlation between $L(\paa)$ and \lir\  \citep[\eg,][]{cal05,alo06}.
For \target\, R08 derived a total IR luminosity in the range
$\lir=5.7-9.5\times 10^{11}$~\lsol, taking  into account the
systematic uncertainty owing to the choice of IR SEDs \citep[see also
the discussion in ][]{pap07}. Here, we reanalyze the total IR
luminosity of \target\ by fitting different sets of IR SEDs to the
flux density at 70, 450, and 850~\micron\ flux densities, all of which
sample the thermal IR emission (see figure~\ref{fig:irsed} and
\S~4.3).  We exclude the 24~\micron\ flux density from this analysis
because it probes $\sim$6--7~\micron\ in the rest--frame mid--IR, and
its relationship to the thermal far--IR emission is not
straightforward.  Figure~\ref{fig:irsed} shows the best fit IR SED
templates and  measured flux densities.   Using the \citet{dal02}
templates, we derive $\lir = (5.0 \pm 0.6) \times 10^{11}$~\lsol, with
a goodness-of-fit, $\chi^2/\nu = 3.2$ and $\nu=2$.  We find formally
better fits using both the IR SEDs of \citet{cha01}, giving $\lir=
(5.9\pm 0.6) \times 10^{11}$~\lsol\ with $\chi^2/\nu = 0.8$, and
\citet{rie09}, giving $\lir = (10 \pm 1.1)\times 10^{11}$~\lsol\ with
$\chi^2/\nu = 0.2$.  The IR luminosities from the fits are consistent
with the results from R08, where the difference here is that we have
excluded the 24~\micron\ flux density and include the deeper
70~\micron\ flux--density measurement.  Because the different IR SED
templates all are consistent with the data, they imply there is a
factor of 2 uncertainty on the total IR luminosity owing to
differences in the choice of template.    Moreover, because the IR
SEDs we have tested are not continuous in $\lir$, this implies there
is \textit{at least} a factor of 2 uncertainty on total IR
luminosities of $z\sim 2.5$ galaxies, even when flux densities are
available at 70~\micron\ and sub-mm wavelengths.

The total IR luminosities correspond to a range of SFR, $\psi =
90-180$~\msol\ yr.   The upper range of the SFR corresponds to the fit
using the Rieke et al.\ template, and these are consistent with the
SFR derived from the \paa\ luminosity.   The templates from Dale \&
Helou and Chary \& Elbaz yield lower \lir\ values, and imply SFRs
lower by $\sim$40--50\% compared to that from $L(\paa)$.  To improve
the accuracy on the IR luminosity will require flux density
measurements at observed wavelengths of $\approx$150--250~\micron, in
order to constrain the peak of the thermal dust emission (see
figure~\ref{fig:irsed}).  Nevertheless, the current analysis provides
evidence that the total IR luminosity and \paa\ luminosities are
consistent for \target\ within a factor of 2. 

Parenthetically, we note that \textit{none} of the empirical IR templates are able simultaneously to fit both the thermal IR emission and the strength of the PAH emission features in the mid--IR.    This is apparent in the lower panel of figure~\ref{fig:irsed} where the best--fit IR templates imply lower observed 24~\micron\ flux densities less than the measured value.  This issue has been identified in the analysis of other  high--redshift gravitationally lensed galaxies \citep[\eg,][]{sia08}.

\subsection{The \paa\ Luminosity Compared to the 8~\micron\ Luminosity}

The origin of the mid--IR emission (rest--frame 5--10~\micron) in
local galaxies is attributed both to very small-grain dust continuum
emission and  molecular PAH emission.  Both the PAHs and very small
grains are heated both by ionizing and non--ionizing sources (in
particular, the ambient galactic radiation field from evolved stars,
Li \& Draine 2002).    Local galaxies and star--forming regions show a
nonlinear relation between  $L(8\micron)$ and $L(\paa)$ \citep[see,
\eg,][]{for04,cal07}, where the slope of the correlation depends on
the dust--heating source(s), the dust spallation/formation rates, and
the metallicity \citep[\eg,][]{hou04b,eng05,cal07,draine07}.  

The MIPS 24~\micron\ flux density probes rest--frame mid--IR,
$\sim$6--7~\micron\ at $z$=2.515, as illustrated in
figure~\ref{fig:irsll}.  Many studies of the IR emission in
distant galaxies use the measured 24~\micron\ flux density
to estimate the total IR emission, and we test these relations for \target.   Using the 24~\micron\ flux density with the prescription of \citet{pap06} yields a estimated total IR luminosity, $\lir = (1.2\pm 0.1) \times 10^{12}$~\lsol, where the error is statistical only and
does not include systematic uncertainties (see discussion in Papovich et al.\ 2006).    This corresponds to a SFR, $\psi \simeq 220$~\msol\ yr$^{-1}$.  While this is consistent with the SFRs derived from the \paa\ emission and \lir\ measured from the far--IR data, this is somewhat fortuitous because the IR template used to
extrapolate the observed 24~\micron\ flux density implies higher flux
densities at observed--frame 70 and 850~\micron\ compared to the
observations.  This is similar to the statement in \S5.1 that \textit{none} of the template IR SEDs are capable to fit simultaneously the far--IR flux densities and mid--IR emission features.   Using the scaling relation from \citet{pap07}, which
includes bolometric corrections using the average 70 and 160~\micron\
flux densities of $1.5 < z < 2.5$ galaxies, yields a nearly equal
estimate for $\lir$ as using the 24~\micron\ data only, with a similar offset compared to the implied SFR from \paa. 

R08 derive a scaling relation between the rest--frame 8~\micron\
luminosity, $L(8\micron)$, and $\lir$ for their sample of
gravitationally lensed $z\sim 2$ galaxies, which includes \target.
Using the $L(8\micron)$ derived in \S~4.3, the R08 scaling
relation yields $\lir = (7.9\pm 1.5) \times 10^{11}$~\lsol.
This is consistent within the range of the total IR luminosity derived
above, and consistent with the implied SFR from \paa\ within the errors.  

\citet{pop08} derive scaling relations between \lir\ and the luminosity of the 6.2 and 7.7~\micron\ PAH emission features.  
%Although R08 measured the 6.2 and 7.7~\micron\ PAH fluxes, the me
%R08 measured the fluxes of the 6.2 and 7.7~\micron\ PAH
%features in \target, and we convert these to luminosities,
%$L(6.2\micron,\mathrm{PAH}) = (4.2\pm0.2)\times 10^{43}$~erg s$^{-1}$
%and $L(7.7\micron,\mathrm{PAH}) = (15.3\pm 0.1)\times 10^{43}$ erg
%s$^{-1}$, corrected for the gravitational lensing magnification.  
R08 measured their PAH luminosities for \target\
simultaneously using PAHFIT (Smith et al.\ 2007).  However, as discussed by Pope et al., line fluxes from measured by PAHFIT are generally higher than those using their method (see also the discussion in Sajina et al.\ 2007; Siana et al.\ 2008).   Thus, we remeasured the 6.2 and 7.7~\micron\ PAH luminosities individually, fitting each emission line with a Drude profile while fitting the slope and intercept of the continuum.   Our fits for each line yielded $L(6.2\micron,\mathrm{PAH}) = (2.25 \pm 0.08)\times 10^{43}$~erg s$^{-1}$ and $L(7.7\micron, \mathrm{PAH}) = (8.89 \pm 0.04)\times 10^{43}$~erg s$^{-1}$.
Using equations 4 and 5
from Pope et al.\ we infer $\lir
\approx 2.3\times 10^{12}$~\lsol\ and $3.1\times 10^{12}$~\lsol\ for
the 6.2 and 7.7~\micron\ PAH features, respectively.   While these estimates of \lir\ are higher by a factor of order three compared to that measured from the far-IR data,  they are within the scatter observed in the the relation between the PAH luminosities and total IR luminosity in the Pope et al.\ sample.
However, the intrinsic IR luminosity of \target\ is a factor of two lower than the high--redshift sub--mm from Pope et al., and it is possible the extrapolated relation does not hold.  This may be supported by the results of \citet{sia08,sia09}, who observe a similar offset between the PAH luminosity and total IR luminosity in their study of intrinsically less-luminous, lensed UV--bright objects. 

Figure~\ref{fig:8um} shows the \paa\ luminosity against the
rest--frame 8~\micron\ luminosity for \target, compared  against
luminosities for samples of local galaxies and  star-forming regions
\citep{cal05,cal07,eng05,dia08}.    \citet{cal07} derive a scaling
relation between the \paa\ and 8~\micron\ luminosities, $L(8\micron)
\propto L(\paa)^{\alpha}$, with $\alpha=0.94$. Note that Calzetti et
al.\ derived this correlation in terms of luminosity \textit{surface
densities} (luminosity per unit physical area).  However,
\citet{dia08} obtain a similar slope for the correlation between the
8~\micron\ and \paa\ \textit{luminosities}.  While the luminosities
for \target\ appear broadly consistent with the extrapolated
relations,  a clearer picture is evident by comparing the ratio of
$L(8\micron)/L(\paa)$, shown in the right panel of
Fig~\ref{fig:8um}.    There is much scatter in the local samples, but
$L(8\micron)/L(\paa)$ for \target\ agrees broadly with those
extrapolated relationships that include the star--forming galaxies
\citep{cal07,dia08}.  Using only the extrapolated relation from
individual \ion{H}{2} regions would underpredict the amount of
8~\micron\ emission \citep[cf.,][]{cal05}.  This implies that  the fraction of
photons from star--formation reradiated as 8~\micron\ luminosity is
weakly dependent on ionizing luminosity.

\subsection{The \paa\ Luminosity Compared to the 24~\micron\ Luminosity}

The emission at rest--frame 24~\micron\ results from thermal dust grains heated by
ionizing and non--ionizing sources.  Empirically, local galaxies and
star-forming regions follow a correlation with $L(24\micron) \propto
L(\paa)^\alpha$, with $\alpha$ in the range 1.03--1.23
\citep{cal05,cal07,alo06}.  \citet{alo06} argue that the correlation
with $\alpha > 1$ arises as dust absorbs ionizing and UV--continuum
photons with increased efficiency in more heavily obscured, more
luminous systems, so that an increasing fraction of the bolometric
luminosity associated with star formation emerges in the IR with
warmer dust temperatures \citep{lon87,wang96,draineli07}.  A similar
conclusion is reached by \citet{cal07}, who argued that the $\alpha >
1$ correlation exists because objects with increasingly higher
starlight intensity and $L(\paa)$ have higher dust
temperatures, where the peak of the emission moves to shorter IR
wavelengths.  \citet{cal07} and \citet{draineli07} discuss the
physical basis for this correlation. 

%The model of Calzetti et al.\ reproduces the trends in their data
%  very  well, including both local galaxies and star--forming regions.
% In addition, observationally, evolved stellar populations contribute
% to the dust heating, and this adds to the IR luminosity
% \citep[\eg,][]{pop00,mis01,bell03}.  Both Alonso--Herrero et al.\ and
% Calzetti et al.\  discuss that the ambient galactic light from
% late--type stars may contribute to the dust heating and the IR
% emission in their samples.  As the integrated emission from galaxies
% includes any contribution from late--type stars,  this effect arguably
% affects the  integrated 24~\micron\ luminosity of starburst  galaxies,
% LIRGs, and ULIRGs more so than observations of individual  \ion{H}{2}
% regions within galaxies.    

Figure~\ref{fig:24um} shows the \paa\ luminosity plotted against the
rest--frame 24~\micron\ luminosity derived above for \target\ and for
local samples \citep{cal05,cal07,eng05,dan05,alo06}.  Interestingly,
\target\ has lower $L(24\micron)$ than local galaxies of
comparable $L(\paa)$, where the offset is  $\approx 0.3-0.5$~dex. This
is more apparent when comparing the ratio of $L(24\micron)/L(\paa)$,
shown in the right panel of figure~\ref{fig:24um}.  Furthermore,
because the 70~\micron\ data for \target\ is blended (see \S~3.2), the
rest--frame 24~\micron\ luminosity may be lower than indicated in the
figure, making the offset between \target\ from the relation for the
local samples more pronounced. 

\target\ has an estimated total IR luminosity,
$L_\mathrm{IR}=(5-10)\times 10^{11}$~\lsol, comparable to local
ultraluminous IR galaxies (ULIRGs, $\lir \ge 10^{12}$~\lsol).
Figure~\ref{fig:24um} includes data for local ULIRGs from the sample
of \citet{dan05}, where we combine their \paa\ measurements with IRAS
25~\micron\ measurements from the literature\footnote{see Moshir,
Kopman, \& Conrow (1992); the NASA Extragalactic Database (NED),
http://nedwww.ipac.caltech.edu/}.  The Dannerbauer et al.\
measurements of \paa\ come from longslit near--IR spectroscopy, and we
have made no attempt to correct for emission outside the slit.  
The \paa-emission in many of local ULIRGs likely results from very compact, nuclear regions, and therefore the spectroscopic slit should contain most of this emission. Nevertheless, we removed from the local ULIRG sample those objects with 2MASS
isophotal diameters $>$15\arcsec, and we also removed those objects
with spectroscopic signatures of AGN.  These steps excluded roughly
one--third of the sample, including the most egregious outliers.
Nevertheless we caution that significant uncertainty may remain due to
primarily unknown \paa\ emission outside the spectroscopic slit or
other aperture effects.  

Even with these caveats the local ULIRGs follow the extrapolation of
the local $L(24\micron)$ and $L(\paa)$ relation, but they show a large
scatter.  We suspect that the large scatter results for the reasons
discussed above, and there may be additional components to the dust
heating beyond ionization from early--type stars, including AGN and
the ambient galactic emission. However, only one galaxy in the local
ULIRG sample has $L(24\micron)/L(\paa)$ and $L(\paa)$ comparable to
\target, and this galaxy (IRAS 04384--4848) has a highly uncertain
dust correction \citep{dan05}.  This implies that no (or at best,
\textit{few}) low redshift ULIRGs have similar physical conditions
producing comparable ratios of mid-IR-to-\paa\  luminosity.

\citet{ken09} combine \ha\ emission--line measurements (uncorrected for dust extinction) and IR continuum measurements of local star--forming galaxies, and derive SFR calibrations of the form, $\psi = 7.9\times 10^{-42} \times [ L(\ha)_\mathrm{obs} + a_\lambda L(\lambda)]$. For the IRAC 8~\micron\ and MIPS 24~\micron\ rest--frame bands they obtain $a_{8} = 0.011$ and $a_{24}=0.020$.   Using the mid--IR luminosities for \target\ derived above, and $L(\ha)_\mathrm{obs} = 3.7 \times 10^{8}$~\lsol\ \citep{kne04}, we obtain $\psi \simeq 45$ and 80~\msol\ yr$^{-1}$, for the 8~\micron\ and 24~\micron\ luminosities, respectively.   These are lower by factors of 4 and 2 compared to that derived from the \paa\ luminosity, but they are within the dispersion reported by \citet{ken09}.  The intrinsic SFR of \target\ is also considerably larger than the objects used to calibrate these relations (see also Moustakas \& Kennicutt 2006), and it is possible the calibrations do not apply under extrapolation.  Larger samples of luminous, high--redshift galaxies are needed to test these relations.

\begin{figure}[t]
\ifsubmode
\epsscale{1.0}  
\else
\epsscale{1.2}
\fi
\plotone{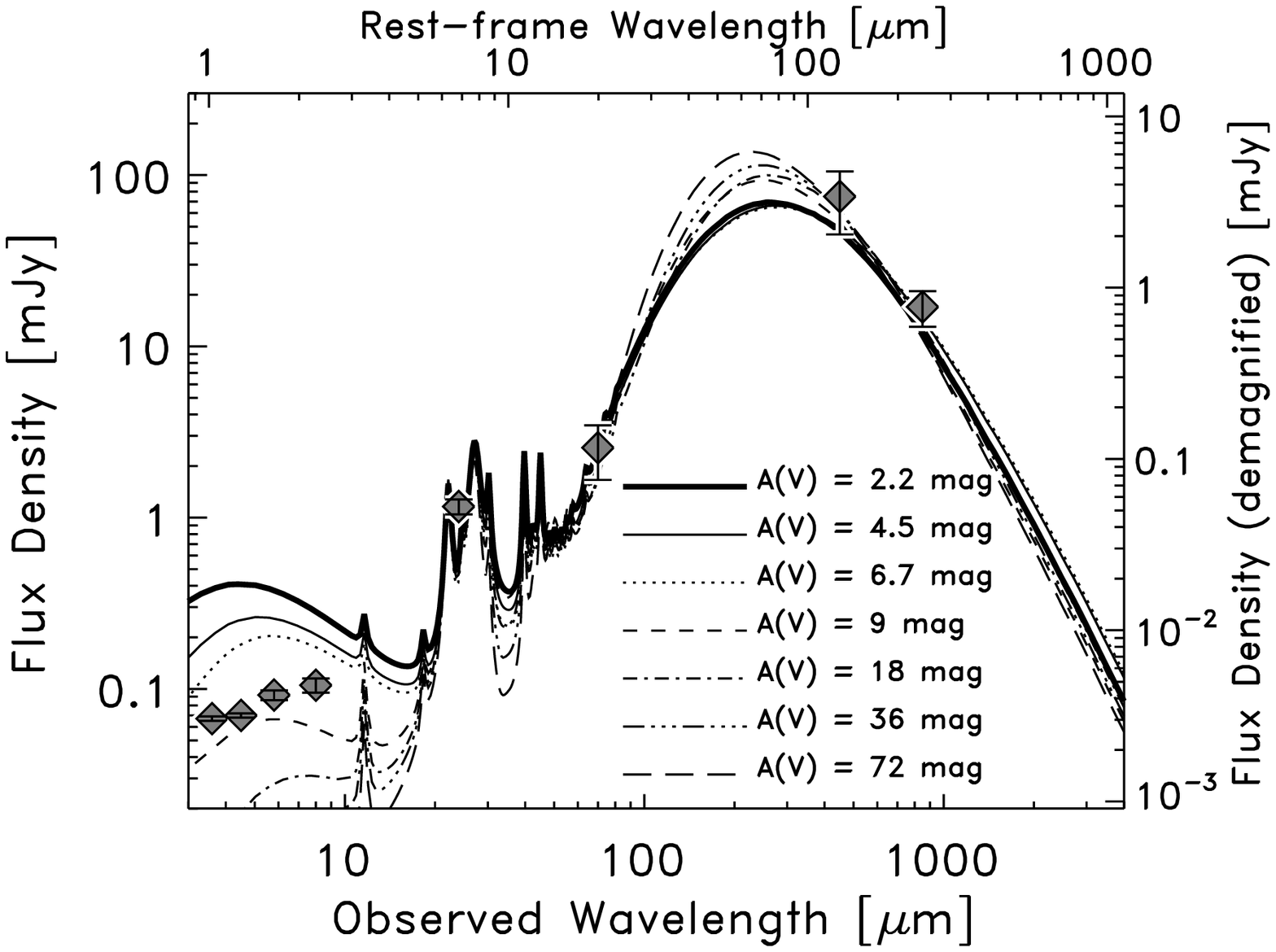}
\epsscale{1.0}  
\caption{ Infrared SED of \target\ compared to radiative transfer models of \citet{sie07}.  The data points show the measured flux densities at IRAC 3.6, 4.5, 5.8 and 8.0~\micron, the MIPS 24 and 70~\micron, and the SCUBA 450 and 850~\micron.  The right axis shows the flux densities after correcting for the gravitational lensing magnification.  The top axis shows the rest--frame wavelength for $z=2.515$.  The curves show model fits to the 24, 70, 450, and 850~\micron\ flux densities.  These SEDs are computed for a spherical PDR ionized uniformly by an interior starburst.  The resulting amount of visual extinction is a variable in the model.  Each curve shows the best--fit model as a function of visual extinction, as indicated in the figure inset.  
\label{fig:irrs07}}
\end{figure}

\subsection{The Nature of \target} 

\target\  appears to host heavily obscured
star--formation at a rate, $\psi \approx 170$~\msol\ yr$^{-1}$.   The
hydrogen ionization rate is $Q_0 = (1.6\pm 0.3) \times
10^{55}$~$\gamma$ s$^{-1}$, implying \target\ may contain as many of
$\sim 10^6$ O stars \citep{stern03}.   Given  the estimate for the
molecular--gas mass ($M[\mathrm{H}_2] \approx 4.5\times 10^9$~\msol,
Kneib et al.\ 2005),  this galaxy could sustain this SFR for $t \sim
30$~Myr.   The starburst in \target\ has had a duration of $\le$100
Myr  based on the analysis of rest--frame UV-to-near-IR SED (see the
Appendix) and supported by the strength of the possible CO index
(\S~4.2).  This is comparable to the dynamical and gas--consumption
timescales based on the observations of the molecular gas.   Because
the gas--consumption timescale is consistent with the dynamical time
and the starburst age, \target\ is likely about midway through this
stage of enhanced star formation.  

The \paa--derived SFR for \target\ is consistent with the SFR implied
by the  total IR luminosity and the rest--frame 8~\micron\ luminosity.
However, the rest--frame 24~\micron\ luminosity is significantly lower
than that expected from the \paa\ luminosity based on the local
relations.    This implies that  \target\ lacks a warm ($\sim$100~K)
thermal dust component typical of local \textit{IRAS}--selected
star--forming galaxies of comparable bolometric luminosity
\citep{lon87,cal00}, which drives the non--linear relationship between
the mid--IR luminosity and the \paa\ luminosity \citep[\eg,][]{cal07}.
Indeed, the thermal dust emission of \target\ (see
figure~\ref{fig:irsed}) peaks near $\sim$100~\micron, and is
consistent with nearly pure emission from dust with temperature, $T_D =
52$~K, and emissivity, $\beta=1.5$.  This is very similar to the dust
emission of photodissociation regions (PDRs) in the vicinity of \ion{H}{2} regions
\citep[\eg,][]{lon87,cal00,chu02}.  In contrast, the IR SEDs for local
galaxies with $\lir$ \textit{comparable to } \target\ have significant
contributions of warm ($\gsim70$~K) dust to the IR emission
\citep[\eg,][]{cha01,dal02,sie07,rie09}.  This is a qualitatively
different than what is observed in \target.  Indeed, the fact that no
galaxies in the local ULIRG sample have $L(24\micron)  / L(\paa)$
values as low as \target\ implies that star formation in 
some high--redshift galaxies with ULIRG luminosities is fundamentally
different. The lower ratio, $L(24\micron) / \lir \simeq 0.1$, for
\target\ implies a lower starlight intensity than that in
local galaxies of comparable bolometric luminosity \citep{draineli07}.
Therefore, while \target\ appears to host a massive starburst that is
similar to local star--forming regions, it is dramatically
``scaled--up'' in luminosity (and presumably SFR).     

This conclusion is supported by a comparison of the mid--IR and
far--IR emission of \target\ to the models of \citet{sie07}, who
calculate the IR emission for PDR--like regions with spherical
symmetry of variable size surrounding \ion{H}{2} regions ionized by a
starburst with a variable fraction of the luminosity coming from OB
associations.   As illustrated in figure~\ref{fig:irrs07}, the
best--fitting models from Siebenmorgen \& Kr\"ugel correspond to an
obscured stellar population where $60-90$\% of the luminosity
originates from OB associations.   Such models require emission from
dust clouds with a range of extinction, $A(V) \approx 2-72$~mag.
Models with visual extinction $A(V) \leq 18$~mag that reproduce the
data have an \textit{intrinsic} luminosity
$\approx10^{10}-10^{11}$~\lsol, and this must be scaled up by a
factors of $\approx$10--100 to match the data, producing total IR
luminosities in agreement with that derived above (\S~5.1).  Scaling
the size of the PDR according, this implies a radius for the
ionization--front of $\sim$3-4~kpc, consistent with the spatial extent
of the molecular CO emission of \target\ \citep[$\sim$3~kpc $\times$
1.5~kpc,][]{kneib05}.  Models with higher \textit{intrinsic} IR
luminosity ($\gg10^{11}$~\lsol) do not reproduce the data.   Models
with higher extinction require larger intrinsic luminosities (as much
as an order of magnitude larger for the model with $A(V) = 72$~mag),
and these provide worse fits to the data (thus, they are less
physical).  Furthermore, models with $A(V) \gsim 30$~mag would be
optically thick to \paa\ photons, and cause significant attenuation of
the mid--IR emission.   While such models are physically motivated in
some cases \citep[\eg,][]{rie09}, if this were the case for \target\,
then the attenuation correction to the \paa\ luminosity would imply a
much larger SFR, which should be substantiated by a larger total IR
luminosity.  The general agreement between the dust--corrected \paa\
luminosity and total IR luminosity (\S5.1) excludes models with very
high extinction.

While the models assume a spherical shell-like dust configuration, in
reality the dust clouds are likely clumpy with some covering fraction,
where the ionizing \ion{H}{2} regions associated with the different OB
associations are in close proximity to the dust clouds and PDRs
\citep{wol90}.  Models factoring in the covering fraction produce a
distribution of dust attenuation \citep[\eg,][]{dop05}, and this is
more consistent with the observations.    Figure~\ref{fig:irrs07}
shows that IRAC 3.6, 4.5, 5.8 and 8.0~\micron\ flux densities for
\target\ are not reproduced simultaneously with the far--IR emission
by any of the best-fitting models.  Thus, spherical symmetry of the
starburst and obscuring PDR seems insufficient to describe both the
direct stellar emission and the far--IR emission for this galaxy.
Therefore, we conclude that star formation in \target\ corresponds to
star--forming regions and starbursts in local galaxies with intrinsic
luminosities of $\lir \approx 10^{10}-10^{11}$~\lsol, but that have been
``scaled--up'' by one--to--two orders of magnitude.   Moreover, this
``scaling--up'' of star formation may be common at high redshift.  For
example, \citet{tac06} argued that the significantly more luminous
sub-mm galaxies at $z\sim 2-3$ resembled ``scaled--up'' and more
gas--rich versions of local ULIRGs.   For \target,  R08 found that the
rest--frame mid--IR spectrum from $6-10$~\micron\ is consistent with
the spectra of local starburst galaxies and inconsistent with the
spectra of local ULIRGs.  Here, our analysis of the far--IR data,
coupled with the SFR implied by the luminosity of the \paa\ line,
leads us to a similar conclusion.

The model we constructed in the Appendix using the observed
rest--frame UV--near-IR SED  consists of a double stellar population.
The stellar population that dominates the bolometric emission (and the
total SFR) in \target\ is very obscured by dust that is optically
thick to UV photons.  The stellar population of the subdominant model
is less obscured, and optically thin to UV photons.   This analysis is
consistent with the scenario discussed above, where the starburst in
\target\ corresponds to OB associations in close proximity to PDRs
with a clumpy distribution.   A similar configuration of multiple
star--forming components with variable dust extinction is observed in
the spatially resolved colors of other galaxies at $z\sim 2-3$, and
especially those with evidence for recent mergers
\citep{pap05,law07}. 

It remains to be seen whether the results discussed here for \target\
are typical of other high--redshift galaxies.  Our full observing
program will produce similar data for eleven gravitationally lensed
galaxies in addition to \target, and these will span a wide range of
mass, optical, and IR properties.  For \target, the baryon (stellar +
gas) mass we derived from the rest-frame-UV--to--near-IR SED and the
dynamical mass from molecular observations \citep{kneib05} both
suggest $M \sim 1.5 \times 10^{10}$~\msol.  This mass is typical of
``$L^\ast$'' UV--selected galaxies at these redshifts
\citep{pap01,sha01}, and such objects likely dominate the SFR density
at this redshift \citep{red08}.    However, the dust obscuration and
IR luminosity are much larger in \target\ compared to the UV--selected
samples, and more typical of sub-mm--selected objects
\citep[\eg][]{blain02,chap05} and IR--luminous $K$--band-selected
objects \citep[\eg,][]{pap06,dad07,wuyts08}.  The fact that the
molecular--gas--emission centroid lies between the UV--components led
\citet{kneib05} to argue that the high SFR in \target\ results from
the merger of two progenitors whose properties are similar to those of
``typical'' UV--selected galaxies.     In this case, we are observing
\target\ during perhaps a short--lived stage of enhanced star
formation.   Our study of the SFR indicators in \target\ provides the
first evidence that the total IR luminosity is proportional to the
number of ionizing photons in these situations.

\vspace{12pt}

\subsection{Final Thoughts}

A potential source of bias is that our observations are sensitive to
the integrated emission from \target, and we are combining datasets
with a wide range of angular resolution.     This is of concern as as
our analysis corresponds to the flux--weighted average of the
star--formation properties of an individual galaxy.   The analysis of
the rest--frame UV--to--near-IR SED (see the Appendix) suggests there
are \textit{at least} two stellar populations, with very different
extinction properties.    Nevertheless, our results probably apply to
the dominant, most luminous star--forming components, which dominate
the nebular emission (\paa) and the IR emission.   Therefore, we
suspect that our results are valid for the average properties of
\target.   We note that similar issues arise in the analysis of the
integrated properties of the starburst galaxies
\citep{eng05,alo06,cal07}, and some caution must be applied then when
comparing these with, for example, the analysis of individual
\ion{H}{2} regions which are resolved in both the narrow--band imaging
of the hydrogen recombination lines and the mid--IR data
\citep{cal05,cal07}.  Achieving resolved observations of the \paa\
emission and far--IR emission in \target\ may be possible with the
\textit{James Webb Space Telescope} and large--aperture sub--mm
facilities.   However, further work will be needed to quantify
possible biases in the integrated emission from galaxies.  

\section{SUMMARY}

We observed the galaxy \target\ at $z=2.515$ with \spitzer\
spectroscopy from 5.8--8.0~\micron.   These data are the deepest
\spitzer/IRS data taken with the SL2 module of any galaxy of which we
are aware.   \target\ is a sub-millimeter--selected infrared-luminous
galaxy, and is lensed gravitationally with a magnification of $\mu=22$
by the rich galaxy cluster Abell 2218.   This galaxy is maintaining a
high rate of star formation based on its IR emission, and it has no
evidence for an AGN in either its rest--frame UV or optical spectra,
nor based on its (lack of) X-ray activity.      We find that the
rest-frame--UV--to--near-IR SED  of \target\ is best represented by
the superposition of a double stellar population  with a varying
amount of dust attenuation, and with a total stellar mass of
$\sim$$10^{10}$~\msol\ (corrected for the gravitational lensing
magnification).  In this model the stellar population that dominates
the star formation is heavily extincted, $A(V) \sim 3$~mag, while the
subdominant stellar population is optically thin to UV photons.

We detected the \paa\ emission line in the \spitzer\ spectrum with a
redshift $z=2.515\pm 0.003$.  The luminosity of the \paa\ emission
line is $L(\paa) = (2.05\pm0.33)\times 10^{42}$~erg~s$^{-1}$,
corrected for the gravitational--lensing magnification, with a
rest-frame equivalent width EW$_0 = 363\pm56$~\AA.   We compared the
\paa\ luminosity to the \ha\ luminosity and derived a nebular
extinction of $A(V) = 3.6\pm 0.4$~mag, although this depends
on the assumed attenuation law and remains a systematic
uncertainty.   This is consistent with that derived from modeling of
the  galaxy's rest--frame UV-to-near-IR SED,
and implies that the dust attenuation affects the ionized gas and
stars uniformly.      The dust--corrected \paa\ luminosity is $L(\paa) =
(2.57\pm 0.43) \times 10^{42}$~erg~s$^{-1}$, corrected for the gravitational lensing magnification.  This corresponds to an
ionization rate of $Q_0 = 1.6 \times 10^{55}$ $\gamma$ s$^{-1}$,
implying \target\ contains on the order of $10^6$ O stars.   Assuming
an IMF with a Salpeter--like slope from 0.1 to 100~\msol\ yields a SFR
$\psi=171\pm28$~\msol\ yr$^{-1}$. 

The \paa--derived SFR agrees with the upper range of SFR implied by
the total IR luminosity. The uncertainty on the total IR luminosity is
a factor of order two, primarily due to the lack of data at
$\sim$100--300~\micron\ (rest--frame $\sim$30--100~\micron).   The
measured $\paa$ and rest--frame 8~\micron\ luminosities are consistent
with the extrapolated relation observed in local galaxies and
star--forming regions.  This implies that both the monochromatic
8~\micron\ and total IR luminosities are dominated by heating from
ongoing star--formation and they are proportional to the number of
ionizing photons.   However, the measured rest--frame 24~\micron\
luminosity is significantly lower in \target\ compared to local
galaxies with comparable \paa\ luminosity.  Thus,  \target\ appears to
lack a warmer dust component ($T_D \sim 70$~K), which is typical in
local galaxies of comparable $L(\paa)$.     The nature of \target\
seems very different from the properties of local ULIRGs, even though
they have comparable SFRs.  Comparing the IR emission of \target\ to
expectations from empirical and radiative transfer models, we conclude
that the starburst in \target\ has similar physics as those in local
galaxies with intrinsic luminosities, $\approx 10^{10}-10^{11}$~\lsol, but
``scaled--up'' by one--to--two orders of magnitude.

The implied stellar and dynamical masses of \target\ are consistent
with those of typical ``$L^\ast$" UV--selected objects, which dominate
the SFR density at this redshift. The implied timescales and starburst
age are on the order of the dynamical time, implying that this galaxy
is only part way through its elevated star--formation episode.   Our
analysis here provides the best measurement yet of the SFR in a galaxy
involved in such an episode and it shows that the IR--luminosity
traces the total SFR in this situation.       While our analysis here
pertains to only one galaxy of any significant redshift, as we extend
this work to our larger sample, it will allow us to study galaxies
spanning a wider range of star--forming properties and luminosities. 

\acknowledgments

The authors acknowledgment invaluable discussions with colleagues that
led to the analysis and interpretation in this paper.   In particular,
we wish thank Almudena Alonso--Herrero, Lee Armus, Miwa Block, Daniela
Calzetti, Sukanya Chakrabarti, Darren Depoy, Vandana Desai, Tony
D{\'i}az-Santos, Naveen Reddy, Johan Richard, Dimitra Rigopoulou, Mark
Swinbank, and Harry Teplitz, for conversations and assistance with the
analysis in this paper.  We also thank the referee, Brian Siana, for
critical comments and suggestions, which improved the paper.  This
research has made use of the NASA/IPAC Extragalactic Database (NED)
which is operated by the Jet Propulsion Laboratory, California
Institute of Technology, under contract with the National Aeronautics
and Space Administration.   This work is based on data obtained with
the Spitzer Space Telescope, which is operated by the Jet Propulsion
Laboratory (JPL), California Institute of Technology (Caltech) under a
contract with NASA.  Support for this work was provided by NASA
through a contract issued by JPL, Caltech under a contract with NASA.
Further support was provided by Texas A\&M University. 

 {\it Facility:} \facility{Spitzer (IRS, MIPS)}

\section*{APPENDIX}

\section*{Stellar Population Fitting}

\begin{figure}[tb]
\ifsubmode
\else
\epsscale{1.2}
\fi
 \plotone{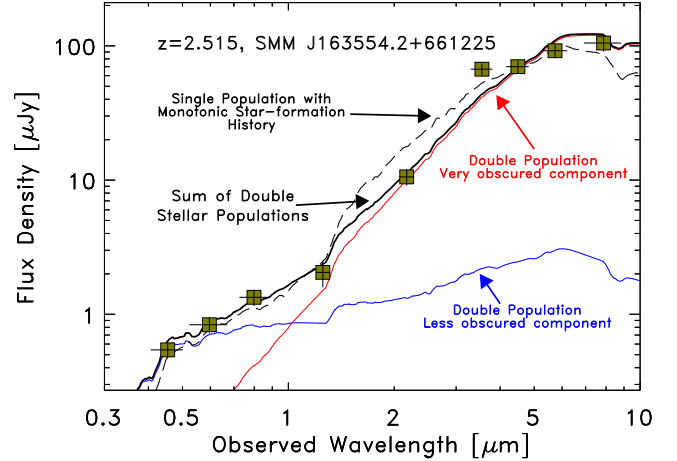}
\epsscale{1.}
 \caption{Spectral energy distribution of \target\ from 0.5 to 8~\micron\ in the observed frame.  The data points show the measured photometry taken from K04 and R08, uncorrected for gravitational lensing.  We find that simple, single--component models with exponentially declining star--formation histories do not reproduce the photometry well.  The dashed line shows the best--fit single--component model, with parameters given in the text.   However, a double stellar--population model with two star--forming components better reproduces the photometry, which is illustrated by the solid curves in the figure.   The red curve corresponds to the ``very obscured component''  and the blue curve to the ``less obscured component'', with model parameters given in the text.  The thick black curve shows the sum of these two components.   }\label{fig:sedfit}
\end{figure}

Photometry is available for \target\ from \hst/WFPC2 ($BVI$),
WHT/INGRID ($J\ks$), and \spitzer/IRAC (channels 1--4), spanning
0.5--8~\micron\ in the observed frame (K04, R08).  We show the SED of
this galaxy in figure~\ref{fig:sedfit}.    We fitted a suite of
stellar population synthesis models to these data to estimate the
properties of the stellar populations in this galaxy.  We used both
the models of \citet{bru03} and the updated 2007 version, which
include a more physically motivated treatment of the thermally
pulsating asymptotic giant branch stars.  We use models with a
Chabrier IMF, although for consistency with
other SFR indicators, we multiply the derived stellar masses and SFRs
by a factor of 1.8 to convert them to equivalent quantities using a
Salpeter-like IMF.  We allow for a range of exponentially declining
star--formation histories with e--folding time $\tau = 1$~Myr
(``instantaneous burst'') to $\tau=100$~Gyr (``constant"
star-formation), metallicities of 0.2 and 1.0 \zsol, and we allow a
range of  dust extinction using the \citet{cal00} law with
$E(B-V)=0-1$.  We note that this dust law may be inappropriate for
heavily obscured sources such as \target\ \citep[see][]{gol02}, but it
provides a useful comparison to other work modeling galaxy spectral
energy distributions (SEDs).   See \citet{pap01,pap06} for details of
the SED fitting.

We find that the photometry are not well fit by simple models with
single component, exponentially declining star--formation histories,
as illustrated in figure~\ref{fig:sedfit}.  Moreover, the best--fit
single--component model is unphysical.  It has a stellar mass of, $M=7
\times 10^{10}$~\msol\ (corrected for the gravitational lensing
magnification), stellar population age, $t=19$~Gyr (greatly exceeding
the age of the Universe at this redshift) , formed with a
star--formation e--folding timescale, $\tau=200$~Myr, and with a color
excess, $E(B-V)=0.32$.  The reduced $\chi^2$ for the best--fitting
single-- component model is $\chi^2/\nu = 14$.   

\target\ likely consists of multiple star--forming components, with
variable extinction and star--formation properties, all of which
contribute  to the rest--frame UV through IR emission.  Such
situations are seen in local IR--luminous galaxies
\citep[\eg,][]{char04}, and arguably apply to high redshift galaxies
as well.  As discussed by K04, \hst\ and ground--based imaging shows
that \target\ consists of several distinct blue components, with a
luminous red core.  We therefore tested SED models consisting of two
star--forming components.  We find that this ``Double Stellar
Population" produces a better fit to the photometry, where the two
components correspond to a very dust--obscured star-forming component,
and a less obscured component.   Figure~\ref{fig:sedfit} shows a
characteristic model that reproduce the data.  The reduced $\chi^2$ of
the fit to the data for the models in the figure is $\chi^2 / \nu =
4.0$, greatly improved over the single--component model above.  In
these models, the ``very obscured component" dominates both the SFR
and the total stellar mass.   In the figure, the very obscured
component consists of a stellar population forming with a constant SFR
with an age, $t=80$~Myr, and a color excess, $E(B-V)=0.80$,
corresponding to $A(V)=3.2$~mag, stellar mass, $M=1 \times
10^{10}$~\msol, and SFR, $\psi=110$~\msol\ yr$^{-1}$ (the stellar mass
and SFR have been corrected for the gravitational lensing
magnification).   The derived color excess is much larger than those
derived for UV--selected galaxies at this redshift
\citep{pap01,sha01}, but is approximately in the range of those
derived for sub--mm selected at this redshift \citep{bor05}.    Also,
the dust extinction we derived from modeling  the  rest--frame
UV-to-near-IR SED here is consistent with that derived from the \ha\
and \paa\ emission--line ratio, and implies that the dust attenuation
affects the ionized gas and stars uniformly.    The subdominant ``less
obscured component" of the double--component model consists of a young
stellar population, $t=40$~Myr, with a star-formation e--folding
timescale, $\tau=50$~Myr, moderate color excess, $E(B-V)=0.30$,
stellar mass, $M=1 \times 10^{8}$~\msol, and SFR, $\psi=3$~\msol\
yr$^{-1}$ (the stellar mass and SFR have been corrected for the
gravitational lensing magnification). The extinction in this component
is consistent with the upper range seen in LBGs at this redshift
\citep{pap01,sha01}.  

Combined with the total gas mass, $M_\mathrm{gas} = 4.5\times
10^{9}$~\msol\ \citep{kneib05}, the total baryonic mass is $M_\ast
\approx 1.5 \times 10^{10}$~\msol, which is consistent with the
estimate of the dynamical mass of the system as inferred from the
molecular gas \citep{kneib05}.       Therefore, we conclude generally
that while the SED of \target\ is complex, it is broadly consistent
with a multiple component starburst, with total stellar mass $M_\ast
\sim 10^{10}$~\msol, where a deeply obscured component is responsible
for for the vast majority ( $\gsim$80\%) of the SFR.

\end{document}